\documentclass[12pt]{iopart}
\pdfoutput=1
\usepackage{color}
\usepackage{amssymb}
\usepackage{graphicx} 

\def\beqra{\begin{eqnarray}}
\def\eeqra{\end{eqnarray}}
\def\beq{\begin{equation}}
\def\eeq{\end{equation}}
\def\etap{\eta^\prime}
\def\etain{\eta_{in}}

\def\vp{\bar{\varphi}}

\def\vp{\varphi}

\def\bx{{\bf{x}}}
\def\bk{{\bf{k}}}
\def\bp{{\bf{p}}}
\def\bq{{\bf{q}}}

\def\bV0{{\bf{V_0}}}
\def\re#1{(\ref{#1})}

\def\agt{~\mbox{\raisebox{-.6ex}{$\stackrel{>}{\sim}$}}~}
\def\alt{~\mbox{\raisebox{-.6ex}{$\stackrel{<}{\sim}$}}~}
\def\bx{{\bf{x}}}

\def\br{{\bf{r}}}

\def\bk{{\bf{k}}}
\def\bp{{\bf{p}}}
\def\bq{{\bf{q}}}

\def\vp{\varphi}

\def\la{~\mbox{\raisebox{-.6ex}{$\stackrel{<}{\sim}$}}~}

\begin{document}
\begin{flushright}
{\small UMN-TH-3604/16}
\end{flushright}

\title[Galilean invariant resummations of cosmological perturbations]{Galilean invariant resummation schemes of cosmological perturbations}

\author{Marco Peloso$^{1}$, Massimo Pietroni$^{2}$}
\vskip 0.3 cm
\address{
$^1$School of Physics and Astronomy, and Minnesota Institute for Astrophysics, University of Minnesota, Minneapolis, 55455, USA\\
$^2$INFN, Sezione di Padova, via Marzolo 8, I-35131, Padova, Italy\\
}

\begin{abstract} 
Many of the methods proposed so far to go beyond Standard Perturbation Theory break invariance under time-dependent boosts (denoted here as extended Galilean Invariance, or GI).  This gives rise to spurious large scale effects which spoil the small scale predictions of these approximation schemes. By using consistency relations we derive fully non-perturbative constraints that GI imposes on correlation functions. We then introduce a method to quantify the amount of GI breaking of a given scheme, and to correct it by properly tailored counterterms. Finally, we formulate resummation schemes which are manifestly GI, discuss their general features, and implement them in the so called Time-Flow, or TRG, equations. 
\end{abstract}

\maketitle

\section{Introduction}
Cosmological perturbation theory (in the following, Standard Perturbation Theory, SPT), as a tool to describe the evolution of structures in the universe has been discussed for a long time \cite{PT,JK06}. It provides a clear improvement with respect to linear theory on the range of scales of the Baryonic Acoustic Oscillations (BAO), but it fails to reach the percent accuracy needed to compare theories and data from present and future surveys. The problem is particularly urgent at small redshifts ($z\alt 2$) and small scales ($k\agt 0.1\;{\mathrm{h/Mpc}}$). Considering higher and higher loop orders in SPT does not lead to an improvement in this regime, as the perturbative expansion seems to converge at most only asymptotically \cite{Blas:2013aba}. The ultimate reason for the SPT failure is its inability to describe short (UV) scales, and their coupling to the intermediate scales relevant for cosmological observations. Numerical studies on the impact of modifications of the initial conditions at UV scales on the late time nonlinear power spectrum (PS) show evidence of a ``screening'' effect which is completely missed by SPT \cite{Little:1991py, 2014arXiv1411.2970N}. The UV failure of SPT is expected, as it neglects from the start all the effects -- such as shell crossing and virialization-- which cannot be described within the pressureless perfect fluid approximation. In order to deal with these UV shortcomings of SPT, modified schemes such as coarse-grained perturbation theory \cite{Pietroni:2011iz, Manzotti:2014loa}, or effective field theory \cite{Baumann:2010tm, Carrasco:2012cv, Blas:2015tla, Floerchinger:2016hja}, have been proposed.

At the opposite end of the spectrum, large scale (IR) modes, are also known to play an important role. Bulk matter flows coherent on $O(10 \, {\mathrm{Mpc/h}})$ scales are responsible for the widening of the BAO peak in the correlation function, and the failure of taking them into account would hinder the use of this feature a standard ruler. A quantity which is very IR sensitive is the cross-correlator of the matter field at different times, or the closely related ``propagator". This quantity is not observable in practice, as observations are confined on the past-light cone, but is nonetheless measurable in simulations by cross-correlating different snapshots and, as we will see, provides powerful tests on the IR performance of a given approximation scheme. 

At each SPT order one can identify  the subset of leading contributions to the  propagator. The perturbative series of this subset is convergent, and can be analytically summed. In the limit in which only very long modes are taken into account, this summation coincides with the Zel'dovich approximation \cite{RPTb}. This observation is at the basis of approximation schemes such as Renormalized Perturbation Theory (RPT) \cite{RPTa}, Multi-Point-Propagator expansion \cite{Bernardeau:2008fa}, or Time-Flow equations \cite{Anselmi:2010fs, Anselmi:2012cn}, in which these leading  IR contributions are summed at all orders, while the SPT expansion for the remaining ones is truncated at some finite order. In what follows, we will refer to all such methods as ``RPT-like summations". Notice that, as will be clear in what follows, also the Lagrangian resummation scheme of \cite{Matsubara07} belongs to this class of methods. 

The problem with RPT-like summations is that they violate Extended Galilean Invariance (GI)  \cite{Turb_Pope},  that is, the invariance with respect to uniform, but time-dependent, boosts of the matter field. A consequence of this symmetry is that the effect of very long IR motions on a equal-time correlation function should vanish, as it can be reabsorbed in a change of frame \cite{Scoccimarro:1995if}. More quantitatively, we will consider a properly defined ``response function" and we will see that IR modes at a wavenumber $q$ should decouple as $O(q^3/k^3)$, where $k\gg q$ is the scale at which the equal time PS is computed.  RPT-like summations fail this test, and generically exhibit a spurious $O(q/k)$ dependence. This is caused by a lack of cancellations between the IR effects that are resummed at all orders and the remaining ones, which are taken into account only up to a fixed order. Once the IR modes are integrated over, the spurious terms typically behave as powers of $k^2 \sigma_{IR}^2$ (where $\sigma_{IR}^2$ is the rms of the displacements induced by the IR modes),  and therefore affect the PS in the UV, namely at large $k$-scales. 

In this paper we address the question of how to quantify the effects of this IR-UV problem in RPT-like summations, how to cure it, and how to formulate summation schemes which are IR safe from the start. Since our primary goal will be  to single out the IR sector while leaving out the UV for a future work, we are not looking for extremely accurate results. We aim at analyzing the problem on general grounds and at identifying common features of the possible solutions. 

We will start with a general discussion on the IR effects on correlation functions at intermediate scales. Using the methods of \cite{Peloso:2013zw},
where exact consistency relations were obtained from GI, we will derive the general structure of the IR effects both for the equal-time and the non-equal time PS. Then we will employ the Zel'dovich approximation as a non-linear -- but analytically manageable -- benchmark to study the emergence of spurious IR effects in RPT like summations, and to envisage possible solutions.  As we will see, a general feature of GI safe resummations is  the dependence on a new scale, which becomes less and less relevant as the order of the truncation is increased. We will introduce a family of resummation schemes parameterised by such scale, which interpolates between SPT on one extreme  and RPT on the opposite one, and we will see that RPT appears as the only pathological (from the point of view of GI) member of the family.
Then, in order to deal with the real dynamics, we will focus on time-flow equations, like the Time Renormalization Group (TRG) of \cite{Pietroni08}, or the ones discussed in \cite{Anselmi:2010fs, Anselmi:2012cn}, clarifying the relation between the two and showing how to introduce the IR effects only partially captured by the TRG in the approximation schemes discussed so far in the literature.

Before closing this introduction, we recall that other GI resummation schemes have been proposed, such as the eRPT of \cite{Anselmi:2012cn, Peloso:2013zw} or the methods of \cite{Senatore:2014via,Baldauf:2015xfa,Blas:2016sfa}. As for the methods discussed in this paper, also these IR-safe resummation introduce an arbitrary scale.

The paper is organised as follows. In Sect.~\ref{formalism} we recall the main equations, introduce our formalism, and clarify the relation between the TRG approach of \cite{Pietroni08} and the time-flow equations of \cite{Anselmi:2010fs, Anselmi:2012cn}, in Sect.~\ref{LRfun} we discuss the linear response function, as the appropriate tool to discuss mode-mode coupling, in Sect.~\ref{ZeldS} we use the Zel'dovich approximation as a benchmark to discuss IR issues and their solutions, in Sect.~\ref{TRGs} we apply the methods developed in the previous section to the real dynamics, using time-flow equations as a tool to resum SPT contributions at all orders. Finally, in Sect.~\ref{conclusions} we give our conclusions.

\section{Equations and formalism}
\label{formalism}

In this section we establish our notation, and summarize the system of equations that we use in this work. For a detailed derivation of these equations we refer the reader to  \cite{MP07b}. We start from the Vlasov equation and the Poisson equation for the dark matter distribution $f \left( \bx ,\, \bp ,\, \tau \right)$ and the gravitational potential. We introduce the moments of the dark matter distribution by integrating over the momentum,~\footnote{Following standard convention, in this and in the following expressions, $\tau$ denotes conformal time, $\bx$ spatial comoving coordinates, $a$ the scale factor of the universe, ${\cal H}$ the Hubble rate in conformal time, $\eta \equiv \log\left( \frac{D_+ \left( \tau \right)}{D_+ \left( \tau_{\rm in} \right)}\right)$ the  logarithm of the linear growth factor.}
\begin{eqnarray} 
\!\!\!\!\!\!\!\! \!\!\!\!\!\!\!\!  \!\!\!\!\!\!\!\! 
n \left( \bx ,\, \tau \right) \equiv \int d^3 p \;  f \left( \bx ,\, \bp ,\, \tau \right) \;\;,\;\;\; 
v^i \left( \bx ,\, \tau \right) \equiv \frac{1}{ n \left( \bx ,\, \tau \right)} \, \int d^3 p \; \frac{p^i}{a \, m} \, f \left( \bx ,\, \bp ,\, \tau \right) \,,\; \dots 
\label{momenta}
\end{eqnarray}
and so on for the higher order. The Vlasov and Poisson equations can be rewritten as an infinite tower of equations for the moments. The tower admits a simple and exact truncation in which only the density field (which we rewrite in terms of the density contrast  $\delta \left( \bx ,\, \tau \right) \equiv \frac{ n  \left( \bx ,\, \tau \right) }{n_0} -1$)  and the velocity divergence   $ \theta  \equiv {\bf \nabla} {\bf  v}$ are non-vanishing. In this paper we will stick to this truncation, also known as ``single stream approximation'', as it is equivalent to neglecting the effect of shell crossing. In \cite{Pietroni:2011iz, Manzotti:2014loa} we developed methods to go beyond this approximation.

 In Fourier space, and in terms of the of the doublet components 
\begin{equation}
\vp_1 \equiv {\rm e}^{-\eta} \,  \delta \;\;,\;\; \vp_2 \equiv {\rm e}^{-\eta} \, \frac{-  \theta}{{\cal H } \, f} \;, 
\label{phi12-def}
\end{equation} 
(where  $f \equiv \frac{1}{\cal H} \, \frac{d \eta}{d \tau}$) the truncation acquires the very compact form 
\begin{equation} 
\!\!\!\!\!\!\!\! \!\!\!\!\!\!\!\!  \!\!\!\!\!\!\!\!  \!\!\!\!\!\!\!\! 
\left( \delta_{ab} \, \partial_\eta + \Omega_{ab} \right) \vp_b \left( \bk ,\, \eta \right) = I_{\bk ,\, \bq_1 ,\, \bq_2} \, {\rm e}^\eta \gamma_{abc} \left( \bq_1 ,\, \bq_2 \right) \vp_b \left( \bq_1 ,\, \eta \right)  \vp_c \left( \bq_2 ,\, \eta \right)  \;, \;\; 
\label{eqs2}
\end{equation} 
where 
\begin{equation}
\!\!\!\!\!\!\!\! \!\!\!\!\!\!\!\!  \!\!\!\!\!\!\!\! 
{\bf \Omega } \equiv \left( \begin{array}{cc} 1 & -1 \\ - \frac{3}{2} \, \frac{\Omega_m}{f^2} &  \frac{3}{2} \, \frac{\Omega_m}{f^2} 
\end{array} \right) \;\;\;,\;\;\;  I_{\bk ,\, \bq_1 ,\, \bq_2} \equiv \int \frac{d^3 q_1}{\left( 2 \pi \right)^3} \,  \frac{d^3 q_2}{\left( 2 \pi \right)^3} \, \left( 2 \pi \right)^3 \delta_D \left( \bk - \bq_1 - \bq_2 \right) \;, \;\; 
\end{equation} 
(with $\delta_D$ being the Dirac delta function),  and where the only non-vanishing components of the vertex functions are
\begin{equation}
\gamma_{112} \left( \bq ,\, \bp \right) = \gamma_{121} \left( \bp ,\, \bq \right) =  \frac{\left( \bp + \bq \right) \cdot \bp}{2 \, p^2} \;\;,\;\; 
\gamma_{222} = \frac{\left( \bp + \bq \right)^2 \, \bp \cdot \bq}{2 \, p^2 \, q^2} \;. 
\end{equation} 

We are interested in correlators of the density and vorticity field. We can express them through the path integral formulation of  \cite{MP07b}. The starting point is the generating functional 
\begin{eqnarray} 
&& \!\!\!\!\!\!\!\! \!\!\!\!\!\!\!\!  \!\!\!\!\!\!\!\!  \!\!\!\!\!\!\!\! 
Z \left[ J_a ,\, K_B \right] \equiv \int {\cal D}  \vp_a  {\cal D}  \chi_b \, {\rm exp} \Bigg\{ - \frac{1}{2} \int d^3 \bk \, \chi_a \left( - \bk ,\, \etain \right) { P}^0 \left( k \right) \, u_a \, u_b \chi_b \left( \bk ,\, \etain \right) + i S \nonumber\\ 
&& \quad\quad + i \int d \eta d^3 \bk \left[  J_a \left( - \bk ,\, \eta \right)   \vp_a  \left(  \bk ,\, \eta \right) +  K_a \left( - \bk ,\, \eta \right)   \chi_a  \left(  \bk ,\, \eta \right) \right]  \Bigg\} \;, 
\label{Z}
\end{eqnarray} 
where $u = \left( \begin{array}{c} 1 \\ 1 \end{array} \right)$ is the growing mode of the linear theory, $P^0 \left( k \right) u_a u_b$ the linear power spectrum  at the initial time $\etain$, which is taken in the infinite past, $\etain\to -\infty$.   $S$ is the action whose extremization gives (\ref{eqs2}): 
\begin{eqnarray} 
&& \!\!\!\!\!\!\!\! \!\!\!\!\!\!\!\!  \!\!\!\!\!\!\!\!  \!\!\!\!\!\!\!\! 
S = \int d \eta \frac{d^3 k}{(2 \pi)^3} \; 
{ \chi}_a \left( - \bk ,\, \eta \right) \Bigg[ 
\int d \etap g_{ab}^{-1} \left( \eta ,\, \etap \right) \vp_b \left( \bk ,\, \eta \right) \nonumber\\ 
&& \qquad\qquad \quad\quad - I_{\bk ,\, \bq_1 ,\, \bq_2} \, {\rm e}^\eta \gamma_{abc} \left( \bq_1 ,\, \bq_2 \right) \vp_b \left( \bq_1 ,\, \eta \right)  \vp_c \left( \bq_2 ,\, \eta \right)  \Bigg] \;. 
\end{eqnarray} 
The quantity $g_{ab}^{-1} \left( \eta ,\, \etap \right) = \delta_D \left( \eta - \etap \right) \left( \delta_{ab} \, \partial_{\etap} + \Omega_{ab} \right)$ 
is the inverse of the linear propagator 
\beqra
 &&
g_{ab} \left( \eta,\,\eta' \right) = \left( B_{ab} + {\rm e}^{-\frac{5}{2} \left( \eta - \etap \right)} A_{ab} \right) \theta \left( \eta - \etap \right) \;,\;\; \nonumber\\
&&
{\bf B} = \frac{1}{5} \left( \begin{array}{cc} 3 & 2 \\ 3 & 2 \end{array} \right) \;,\;\; 
{\bf A} = \frac{1}{5} \left( \begin{array}{cc} 2 & - 2 \\ - 3 &  3 \end{array} \right)  \;, 
\label{g-AB}
\eeqra
with $\theta$ is the Heaviside step function 

Second derivatives of $W \equiv - i \, \ln \, Z$ with respect to the sources $J_a$ and $K_a$ provide the nonlinear power spectrum 
\begin{eqnarray} 
{ P}_{ab} \left( k; \eta ,\, \eta' \right) (2 \pi)^3\delta^{(D)} \left( \bk + \bk' \right) && \equiv   \left\langle \vp_a \left( \bk ,\, \eta \right) \,  \vp \left( \bk' ,\, \eta' \right) \right\rangle \nonumber\\
   &&=   -i \, \frac{\delta^2 W}{\delta J_a \left( - \bk ,\, \eta \right) \delta J_b \left( - \bk' ,\, \etap \right) } \Big\vert_{J_a = K_a = 0}\,, \nonumber\\ 
\label{P-def} 
\end{eqnarray} 
and propagator 
\begin{eqnarray}
G_{ab} \left( k; \eta ,\, \eta' \right)  (2 \pi)^3  \delta^{(D)} \left( \bk + \bk' \right) &&\equiv  - i \left\langle \vp_a \left( \bk ,\, \eta \right) \,  \chi_b \left( \bk' ,\, \eta' \right) \right \rangle\nonumber\\
&& =  -  \frac{\delta^2 W}{\delta J_a \left( - \bk ,\, \eta \right) \delta K_b \left( - \bk' ,\, \etap \right) } \Big\vert_{J_a = K_a = 0} \,.\nonumber\\ 
\label{G-def} 
\end{eqnarray}

In these expressions  brackets denote, as usual, averaging over the initial conditions at time $\etain$. In this average, the initial conditions obey gaussian statistics, due to the $\chi^2 P^0$ factor in eq. (\ref{Z})  \cite{MP07b}. A more general and non-gaussian initial statistics can be imposed on the system  by adding terms cubic, quartic, etc., in $\chi(\etain)$, with coefficients given by the initial bispectrum, trispectrum, and so on. 

As customary in field theory, we then define  \cite{MP07b}  the expectation values of the $\vp_a$  and $\chi_b$ in the presence of sources, 
\begin{eqnarray} 
\vp_a \left[ J_c , K_d \right] \equiv \frac{\partial W \left[ J_c , K_d \right]}{\partial J_a} \;\;,\;\; 
\chi_b \left[ J_c , K_d \right] \equiv \frac{\partial W \left[ J_c , K_d \right]}{\partial K_b} \;, 
\label{phi-chi-exp}
\end{eqnarray}
and the generating functional for one particle irreducible Green functions
\begin{equation}
\Gamma \left[ \vp_a ,\, \chi_b \right] \equiv W \left[ J_a ,\, K_b \right] - \int d^\eta \frac{d^3 k}{(2 \pi)^3} \left( J_a \, \vp_a  + K_b \, \chi_a \right)  \;.   
\label{gamma}
\end{equation} 

This is used in the definition of the ``self-energy'' and of the mode coupling function  
\begin{eqnarray}
 \!\!\!\!\!\!\!\! \!\!\!\!\!\!\!\!  \!\!\!\!\!\!\!\!  \!\!\!\!\!\!\!\! 
\frac{\partial^2 \Gamma}{\partial  \vp_a \left( \bk ,\, \eta \right) \, \partial  \vp_b \left( \bk' ,\, \etap \right) } \Big\vert_{ \vp_a =  \chi_b = 0}' 
&\equiv&  g_{ba}^{-1} \left( \etap ,\, \eta \right) - \Sigma_{ba} \left( k ; \etap ,\, \eta \right)  \;\;, \nonumber\\ 
 \!\!\!\!\!\!\!\! \!\!\!\!\!\!\!\!  \!\!\!\!\!\!\!\!  \!\!\!\!\!\!\!\! 
\frac{\partial^2 \Gamma}{\partial  \vp_a \left( \bk ,\, \eta \right) \, \partial  \chi_b \left( \bk' ,\, \etap \right) } \Big\vert_{ \vp_a =  \chi_b = 0}' 
&\equiv& i  
{ P}^0 \left( k \right) u_a u_b \delta_D \left( \eta-\etain \right) \delta_D \left( \eta'-\etain \right) + \Phi_{ab} \left( k ; \eta ,\, \eta' \right)   \;\;. \nonumber
\label{d2Gamma} 
\end{eqnarray}

Starting from the definitions (\ref{phi-chi-exp}) and (\ref{gamma}), one can show that the matrix formed by the second derivatives of $W \left[  J_c , K_d \right]$ and that formed by the second derivatives of $\Gamma \left[ \vp_a ,\, \chi_b \right] $ are minus the inverse of each other  \cite{MP07b}. 
This generates some identities between these elements. One of the identities is 
\begin{eqnarray}
&&  \!\!\!\!\!\!\!\! \!\!\!\!\!\!\!\!  \!\!\!\!\!\!\!\!  \!\!\!\!\!\!\!\!
P_{ab} \left(  k ;\, \eta ,\, \eta' \right) = { P}^p \left(  k ;\, \eta ,\, \eta' \right) +  { P}^{MC} \left(  k ;\, \eta ,\, \eta' \right) \nonumber\\ 
&&
= G_{ac} \left( k; \eta ,\, -\infty \right)  G_{bd} \left( k; \eta ,\, -\infty \right) u_c u_d { P}^{(0)}(k)  \nonumber\\
&&
+\int_{-\infty}^\eta d s  \int_{-\infty}^{\etap} d s'  G_{ac} \left( k; \eta ,\, s \right)  G_{bd} \left( k; \eta ,\, s' \right) \Phi_{cd} \left( k ; s ,\, s' \right) \;, 
\label{Pp-PMC} 
\end{eqnarray}
which separates the full power spectrum in the sum of a ``propagator'' and of a ``mode coupling'' part. As anticipated, in the equation above we have sent $\etain\to \infty$. 

Applying the equation of motion, eq. (\ref{eqs2}), on the definition of the (unequal time) PS,  eq.~\re{P-def}, we obtain the following equation 
\begin{eqnarray} 
&& \!\!\!\!\!\!\!\! \!\!\!\!\!\!\!\!  \!\!\!\!\!\!\!\!   
\partial_\eta { P}_{ab} \left( k ;\, \eta ,\, \eta' \right) =  - \Omega_{ac} \, { P}_{cb} \left( k;\, \eta ,\, \eta' \right)  \nonumber\\ 
&&\qquad \quad+ 
{\rm e}^\eta \, I_{\bk ,\, \bq_1 ,\, \bq_2} \, \gamma_{acd} \left( \bq_1 ,\, \bq_2 \right) \,  \langle   \vp_c \left( \bq_1  ,\, \eta \right)  \vp_d \left( \bq_2 ,\, \eta \right)  \vp_b \left(  -\bq_1-\bq_2 ,\, \eta' \right)   \rangle'  \;, \nonumber\\
\label{TRGPS}
\end{eqnarray}
where prime on a correlator denotes the correlator  divided by $(2 \pi)^3 \delta_D(0)$, the overall momentum delta function. 
This equation is the starting point of the TRG method, introduced in \cite{Pietroni08}. In order to solve it, the bispectrum at the RHS must be computed. In \cite{Pietroni08} this was accomplished by applying  eq.~(\ref{eqs2}) to the bispectrum, in order to obtain  an evolution equation for the latter. This, in turn, involves the trispectrum, and so on. In \cite{Pietroni08} the tower of equations was truncated by setting the trispectrum to zero. The effect of including the (tree-level) trispectrum was studied in \cite{2013MNRAS.428.3173J}.  The system of equations obtained in this way contains perturbative contributions at all orders, and in this sense it provides a resummation of the SPT expansion. However, it fails to properly include the effects of long IR modes, as is clearly manifest if one consider the truncated set of equations for the non-equal time PS.   An alternative to using a (truncated) set of equations to compute the bispectrum is to use the identity (56) of  \cite{Peloso:2013zw} for the full bispectrum, which allows to rewrite eq.~\re{TRGPS} as in \cite{Anselmi:2012cn},
\beqra
&& \!\!\!\!\!\!\!\!\!\!\!\!\!\!  \!\!\!\!\!\!\!\!  \!\!\!\!\!\!\!\! 
\partial_\eta P_{ab}(k;\eta,\etap) =-\Omega_{ac}P_{cb}(k;\eta,\etap)  \nonumber\\ 
&&  +\int_{-\infty}^\eta ds\; \Sigma_{ac}(k;\eta,s) P_{cb}(k;s,\etap)  +\int_{-\infty}^{\etap} ds\;  \Phi_{ac}(k;\eta,s) G_{bc}(k;\etap,s) \,.
\label{eq-PS-neq-exact}
\eeqra
At this stage, the equation is exact, as it is eq.~\re{TRGPS}. In order to solve it, some approximation on the exact two-point 1PI functions $\Sigma_{ab}$ and $\Phi_{ab}$ has to be performed. As we will see, the form \re{eq-PS-neq-exact} provides a more direct path to the resummation of IR effects. 

When solving the equation numerically, initial conditions  are given at some time $\eta_s$, where SPT can be reliably used for all scales of interest. Notice that the initial time $\eta_s$ needs not be the same as $\etain$, the time at which the initial conditions on the perturbations are given, which can be thought as being infinitely far in the past.

We will discuss various approximations of this equation in the remainder of this work, concentrating on how to incorporate IR effects beyond SPT without spoiling Galilean Invariance.

\section{Linear Response function}
\label{LRfun} 

Assuming gaussian initial conditions on the growing mode, and fixing the cosmology, the nonlinear PS at late times,  $P_{ab}(k;\eta,\etap)$, is a functional of the linear PS, given at some initial time $\etain$, $P^0(q;\etain)$. In the following, to simplify the notation, we will omit the $\etain$ dependence when redundant.

We are interested in the response of the nonlinear PS to ``small'' variations of the initial PS. In full generality, we can write the resulting nonlinear PS as a functional expansion,
\beqra
\!\!\!\!\!\!\!\!\!\! \!\!\!\!\!\!\!\!\!\! \!\!\!\!\!\!\!\!\!\! \!\!\!\! P_{ab}[P^0](k;\eta,\etap) = P_{ab}[\bar P^0](k;\eta,\etap)  \nonumber\\
&& \!\!\!\!\!\!\!\!\!\! \!\!\!\!\!\!\!\!\!\! \!\!\!\!\!\!\!\!\!\! \!\!\!\! \!\!\!\!\!\!\!\!\!\!\!\!\!\!\!\! +\sum_{n=1}^{\infty} \frac{1}{n!} \int d^3 q_1\cdots d^3 q_n\; \left.\frac{\delta^n P_{ab}[P^0](\bk;\eta,\etap) }{\delta P^0(\bq_1)\cdots \delta P^0(\bq_n) }\right|_{P^0=\bar P^0} \; \delta P^0(q_1)\cdots \delta P^0(q_n)\,,\nonumber\\
&&\!\!\!\!\!\!\!\! \!\!\!\!\!\!\!\!\!\! \!\!\!\!\!\!\!\!\!\! \!\!\!\!\!\!\!\!\!\! \!\!\!\! \!\!\!\!\!\!\!\! = P_{ab}[\bar P^0](k;\eta,\etap)  + \int \frac{d q}{q}\; K_{ab}(k,q;\eta,\etap) \, \delta P^0(q) + \cdots\,,
\label{NLexp}
\eeqra
where $\delta P^0(q) \equiv P^0(q)-\bar P^0(q)$ is the deviation from the reference linear  PS, $\bar P^0(q)$. Notice that for $\bar P^0(q)=0$ the expansion above is just the SPT expansion, where the the $n$-th term in the series corresponds the $(n-1)$-th loop order. On the other hand, for any $\bar P^0(q)\neq 0 $, already the linear response function (LRF),
\beq
 K_{ab}(k,q;\eta,\etap) \equiv q^3 \int d \Omega_{\bq} \;\left.\frac{\delta P_{ab}[P^0](\bk;\eta,\etap) }{\delta P^0(\bq)}\right|_{P^0=\bar P^0}\,,
 \label{lrf}
\eeq
contains SPT contributions at all orders, that is, arbitrarily high powers in $\bar P^0$, and is therefore a fully nonperturbative object.
 In \re{lrf} we have used spatial isotropy. 

The knowledge of the LRF can be used to obtain the nonlinear PS for a cosmology with a linear PS not too different from the reference one, once the nonlinear PS for the latter has been computed, {\it e.g.}, by N-body simulations. Besides this practical use, the LRF is also relevant for a more fundamental issue, namely, it quantifies, at a fully nonlinear level, the coupling between different modes. More precisely, if we consider a $\delta P^0(q)$ peaked around a fixed momentum $q$, and rapidly vanishing far from it, the LRF encodes how much a (small) modification of the initial condition at a scale $q$ impacts on the nonlinear PS at later times at a scale $k$. 

In ref. \cite{2014arXiv1411.2970N} the LRF was  measured in N-body simulations and the results were compared to SPT.  Strong deviations between SPT and simulations were found in the UV sector ($q/k\ \gg 1$). In particular, while PT predicts a non vanishing, or even growing LRF at large $q$ (for fixed $k$), N-body simulations find quite an opposite behaviour, with the LRF going to zero, thus showing evidence for a decoupling between UV and intermediate scales. A similar ``UV screening"  was already pointed out in \cite{Little:1991py}.  
While this is a very interesting results, which clarifies the reasons of the failure of SPT, in this paper we focus on the opposite, IR regime. In the following, we derive the constraints on the  IR behavior of the LRF coming from the underlying symmetry of the system, namely the GI recalled in the Introduction. 

As a first step, since we are looking for exact statements deriving from GI, we want to express the LRF in terms of the exact propagator, PS, and other $n$-point functions. We start from the definition of the nonlinear PS and its path integral formulation (\ref{P-def}). Although that expression assumes gaussian initial conditions, after eq.~(\ref{G-def}) we have discussed how to generalize it to more general initial conditions. 
The expression for the LRF given in eq.~\re{lrffull} below would not change in this case.

By taking the functional derivative of \re{P-def} with respect to the initial PS we get
\beqra
&& \!\!\!\!\!\!\!\!\!\!\!\! \!\!\!\!\!\!\!\!\!\!\!\!  \!\!\!\!\!\!\!\!\! 
(2 \pi)^3 \delta_D(\bk+\bk') \frac{\delta P_{ab}(\bk;\eta,\etap)}{\delta P^0(\bq)} =  -\frac{1}{2} \frac{1}{(2\pi)^3} \langle  \vp_a(\bk;\eta)  \chi_c(-\bq;\etain)  \chi_d(\bq;\etain) \vp_b(\bk';\etap)  \rangle  u_c u_d\,,\nonumber\\
&&\qquad\qquad\quad\quad\;\;=  \frac{1}{2} \frac{1}{(2\pi)^3} \frac{1}{P^0(q)^2}\langle  \vp_a(\bk;\eta)  \vp^{in}(-\bq)  \vp^{in}(\bq) \vp_b(\bk';\etap)  \rangle \,,
\label{dp}
\eeqra
where we have used the relation, valid when the initial conditions are gaussian and on the growing mode ($\vp_c(\bq;\etain) =u_c\, \vp^{in}(\bq) $),
\beq
 \chi_d(\bq;\etain) u_d = \frac{i}{ P^0(q)} \,\vp(\bq;\etain) .
 \eeq
The case of non gaussian initial conditions leads to  a new $O(q)$ term besides the one in eq.~\re{lrffullIR} below, which, using the methods developed  in \cite{Peloso:2013spa}, can be computed at all orders in SPT but at linear order in the nongaussianity parameter, $f_{NL}$.

\begin{figure}
\centering{ 
\includegraphics[width=0.9\textwidth,clip]{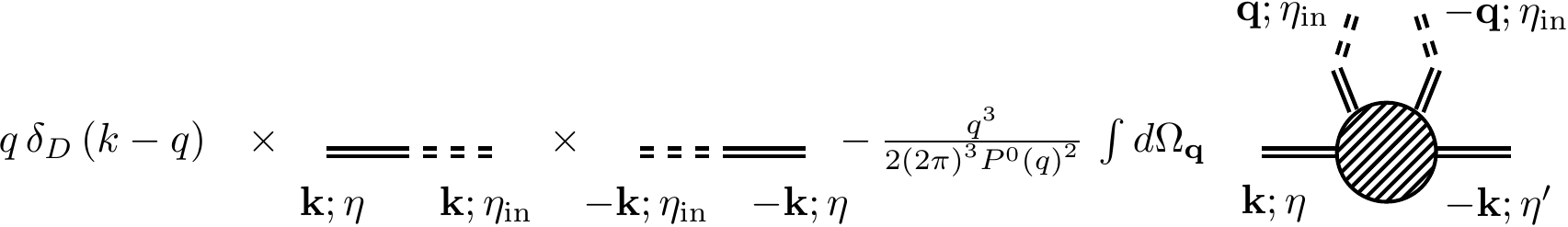}
}
\caption{Diagrammatical representation of the linear response function $K_{ab}(k,q;\eta,\etap)$, eq.~\re{lrffull}. 
The double mixed solid-dashed line denote full propagators, and the dark blob represents the full trispectrum. 
}
\label{lrfgraph}
\end{figure}

Inserting \re{dp} in \re{lrf}, gives
\beqra
&& \!\!\!\!\!\!\!\!\!  \!\!\!\!  \!\!\! K_{ab}(k,q;\eta,\etap) = q\, \delta_D(k-q) \,G_{ac}(k;\eta,\etain)u_c\,G_{bd}(k;\etap,\etain)u_d \nonumber\\
&& \qquad + \frac{1}{2} \frac{1}{P^0(q)^2} \frac{q^3}{(2 \pi)^3} \int d\Omega_{\bq}\,\langle  \vp_a(\bk;\eta)  \vp^{in}(-\bq)  \vp^{in}(\bq) \vp_b(-\bk;\etap)  \rangle_c^\prime\,,
\label{lrffull}
\eeqra
where the first line represents the disconnected contribution to the four-point function in \re{dp}, $\langle  \cdots \rangle_c^\prime$ indicates the connected contribution, and we used the relation
\beq
q^3 \int d\Omega_\bq\,  \delta_D(\bk-\bq) = q\,\delta_D(k-q)\,.
\eeq
Eq.~\re{lrffull} is represented diagrammatically in Fig.~\ref{lrfgraph}.

The consistency relations derived in \cite{Peloso:2013zw, Kehagias:2013yd,Creminelli:2013mca} relate $N+1$-point functions with one of the momenta going to zero with $N$-point ones. Applying them twice to the 4-point function in \re{lrffull} we get
\beqra
&&\!\!\!\!\!\!\!\!\!\!\!\!\!\!  \!\!\!\!\!\!\!\!\!\!\!\!\!\! \!\!\!\!\!\!\!\!   \frac{1}{P^0(q)^2 }\langle  \vp_a(\bk;\eta)  \vp^{in}(-\bq)  \vp^{in}(\bq) \vp_b(-\bk;\etap)  \rangle_c^\prime  \nonumber\\
&&\to -\left(\frac{\bq\cdot\bk }{q^2}\right)^2 \left(e^\eta-e^{\etap}\right)^2  \,P_{ab}(k;\eta,\etap) + O(q^0)\,,
\eeqra
for $q\ll k$. Notice that this relation is exact, in the sense that not only it holds at all orders in SPT but it also holds beyond the perfect fluid approximation, that is, if multi streaming is present.
Inserting it in \re{lrffull} and performing the angular integration, we get the IR limit of the linear response function
\beqra
&& \!\!\!\!\!\!\!\!\!  \!\!\!\!  \!\!\! K_{ab}(k,q;\eta,\etap) \to  -  \frac{1}{3} \frac{1}{(2 \pi)^2}  k^2 q \,\left(e^\eta-e^{\etap}\right)^2\, P_{ab}(k;\eta,\etap) + O(q^3)\,,
\label{lrffullIR}
\eeqra
where we have used the fact that in the IR limit the full propagator gives the linear one $G_{ac}(q;\eta,\etain)u_c \to g_{ac}(\eta,\etain)u_c = u_a$. We stress that GI protects the structure of the $O(q)$ term in \re{lrffullIR} at all orders. On the other hand, it provides no constraint on the coefficient of the $O(q^3)$ terms. Moreover, in the $\eta=\etap$ case, the $O(q)$ term vanishes, as in the consistency relations for the LSS \cite{Peloso:2013spa} \footnote{ Consistency relations among equal time correlators involving momentum instead of density fields have been recently discussed in \cite{Rizzo:2016akm}. }.

\begin{figure}
\centering{ 
\includegraphics[width=1\textwidth,clip]{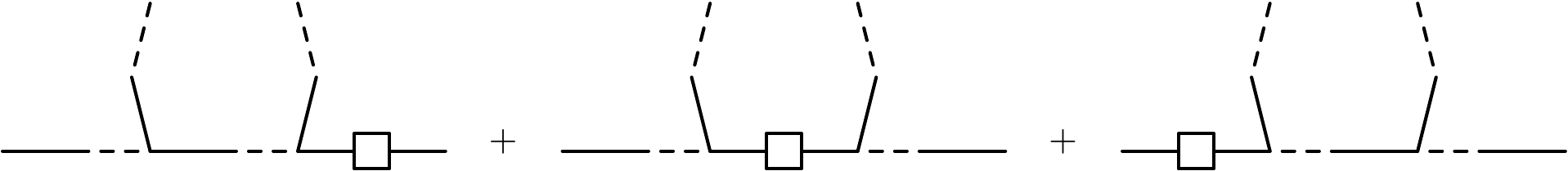}
}
\caption{Diagams contributing to the lowest order PT result for the LRF. Another set of contributing diagrams, in which the vertical lines cross, is not shown. A solid line with a box  (resp., a mixed solid-dashed line) denotes a linear power spectrum (resp., a linear propagator). 
}
\label{lrfPTgraph}
\end{figure}

In SPT, the lowest order contributions to the second term in \re{lrffull} are given by  
the diagrams in Fig.~\ref{lrfPTgraph}, which give, for the density-density component in the IR limit,

\beqra
&&  \!\!\!\!\!\!\!\!\!  \!\!\!\!  \!\!\!\!\!\!\!\!\!  \!\!\!\!  \!\!\!\!\!\!\!\!\!  
K_{11}^{PT}(k,q;\eta,\etap) \to  -  \frac{1}{3} \frac{1}{(2 \pi)^2}  k^2 q \,\left(e^\eta-e^{\etap}\right)^2\, P^0(k) \nonumber\\
&&  \!\!\!\!\!\!\!\!\!  \!\!\!\!  \!\!\!\!\!\!\!\!\!  \!\!\!\!  \!\!\!\!\!\!\!\!\! 
+  \frac{
\left[ 812  \left( {\rm e}^{\eta} + {\rm e}^{\etap} \right)^2 + 1790 \, {\rm e}^{\eta+\etap} \right]  \,P^0(k) 
+ {\rm e}^{\eta+\etap} \left[ -1974 \,k \,\frac{d P^0(k)}{d k} +441\, k^2\,\frac{d^2 P^0(k)}{d k^2} \right] }{8820 \pi^2} \,q^3\,\nonumber\\
&&  \!\!\!\!\!\!\!\!\!  \!\!\!\!  \!\!\!\!\!\!\!\!\!  \!\!\!\!  \!\!\!\!\!\!\!\!\! 
+ O(q^5) \,. \nonumber\\ 
 \label{irlimit}
\eeqra

Notice the difference between the $O(q)$ terms in the full result and in lowest order SPT: while the former contains the fully  nonlinear PS $P_{ab}(k;\eta,\etap) $, the latter contains, obviously, the linear one $P^0(k)$. The approximate fulfilment of the GI requirement provided by SPT is the reason for its imperfect description of physical effects where GI plays an essential role, such as the damping of memory effects encoded in the nonlinear propagator or, on a more observational side, the widening of the BAO peak \cite{Peloso:2015jua}. The cancellation of the spurious IR dependence for the equal-time correlators order by order in SPT was discussed in \cite{Jain:1995kx,Scoccimarro:1995if,Peloso:2013zw,Blas:2013bpa}.

The $q$ dependence  of the IR limit of the LRF  is a practical test for approximation schemes other than SPT. In the equal-time case, it ensures that these schemes  do not bear spurious $O(k^2 q P^0(k))$ effects.

\section{Zel'dovich approximation}
\label{ZeldS}

In this section we consider the Zel'dovich approximation as a benchmark to study the issue of IR dependence and how we can deal with it 
 in resummation schemes. The starting point is to notice that the unequal time density-density PS (in which the density field has been rescaled by $e^{-\eta}$ as in eq. (\ref{phi12-def}))  can be written as
\beq
P_{11}(k;\eta,\etap)= e^{-(\eta+\etap)}\int d^3r_L\; e^{-i\,\bk\cdot\br_L} \left(\langle e^{-i \bk\cdot\Delta{\bf \Psi}} \rangle -1 \right)\,,
\eeq
where $\Delta{\bf \Psi}$ is the difference between the displacement fields,
\beq
\Delta{\bf \Psi} = {\bf \Psi}(\bq_L,\eta)-{\bf \Psi}(\bq'_L,\etap)\,,\qquad\qquad \br_L=\bq_L-\bq_L',
\eeq
and $\br_L$, $\bq_L$ are vectors in Lagrangian configuration space.

The Zel'dovich approximation consists in evaluating the displacement field in linear perturbation theory (growing mode)
\beqra
&& {\bf \Psi}_Z(\bq_L,\eta) =\int^{\tau(\eta)}_0\, d\tau'' {\bf v}(\bq_L,\tau'') = i\,e^\eta\,\int \frac{d^3k}{(2\pi)^3} \,e^{i\,\bq_L\cdot\bk}\; \frac{\bk}{k^2}\, \vp^{in}(\bk),
\eeqra
where the initial density field, $\vp^{\rm in} \left( \bk \right) = {\rm e}^{-\etain} \, \delta^{\rm in}(\bk)$, is assumed to obey gaussian statistics. 
Therefore
\beq
\langle e^{-i \bk\cdot\Delta{\bf \Psi}_Z} \rangle = e^{-\frac{1}{2}k^ik^j \langle \Delta{\Psi^i_Z}\Delta{ \Psi^j_Z} \rangle},
\eeq
which, after some manipulation gives the Zel'dovich PS as
\beqra
&& \!\!\!\!\!\!\!\!\!  \!\!\!\!  \!\!\!\!\!\!\!\!\!  \!\!\!\! \!\!\!\! \!\!\!\! P_Z(k;\eta,\etap)=e^{-(\eta+\etap)} \int d^3r_L e^{-i\,\bk\cdot\br_L}\left[e^{-\frac{k^2 \sigma^2}{2}(e^{\eta}-e^{\etap})^2} \,e^{-e^{\eta+\etap}\,\left(k^2\sigma^2 -I(\bk,\br_L)\right)}-1\right]\,,
\label{PZ}
\eeqra
where 
\beq
I(\bk,\br_L) = \int\frac{d^3 p }{(2\pi)^3}\;e^{i\,\bp\cdot\br_L}\;\frac{(\bp\cdot\bk)^2}{p^4}\,P^0(p)\,,\qquad\qquad  I(\bk,0) = k^2 \sigma^2\,,
\label{I-sigma-def}
\eeq
with
\beq
\sigma^2\equiv \frac{1}{3}  \int\frac{d^3 p }{(2\pi)^3} \frac{P^0(p)}{p^2}\,.
\eeq
Noticing that the $P^0(q)$ dependence in \re{PZ} is contained in the $\sigma$'s and in the $I(\bk,\br_L)$ function at the exponent,  we can compute the functional derivative with respect to the linear PS to get the LRF defined in \re{lrf}, 
\beqra
&&  \!\!\!\!\!\!\!\!\!  \!\!\!\!  \!\!\!\!\!\!\!\!\!  \!\!\!\! \!\!\!\! \!\!\!\! K_{11}^{Z}(k,q;\eta,\etap)= q\, \delta_D(k-q) G_z(k;\eta)G_z(k;\etap) -  \frac{1}{3} \frac{1}{(2 \pi)^2}  k^2 q \,\left(e^\eta-e^{\etap}\right)^2\, P^Z(k;\eta,\etap) \nonumber\\
&&\qquad -  \frac{q^3}{(2\pi)^3} G_z(k;\eta)G_z(k;\etap) \int d\Omega_\bq \frac{(\bq\cdot\bk)^2}{q^4} {\cal J}(\bk,\bq)\,,\nonumber\\
\label{LRFZ}
\eeqra
where
\beq
G_z(k;\eta)=e^{-\frac{k^2\sigma^2}{2}e^{2\eta}}\,,
\eeq
and
\beq
  {\cal J}(\bk,\bq) \equiv \int d^3 r_L\, \; e^{-i\,\bk\cdot\br_L}  \left(1-\cos\left(\bq\cdot\br_L \right)\right) \left(e^{e^{\eta+\etap}I(\bk,\br_L)}-1 \right) \,.
\eeq
Considering eq.~\re{LRFZ} in the $q\ll k$ limit, we are left with the second term, which has the same structure as the exact result \re{lrffullIR}, and the third term, which is $O(q^3)$ as expected. Notice that, compared to 1-loop SPT, the $O(q)$ contribution is recovered by the Zel'dovich approximation at the fully nonlinear level.

The absence of spurious $O(q)$ terms from the second line is due to the presence of the combination $\left(1-\cos\left(\bq\cdot\br_L \right)\right)=O(q^2 r_L^2)$, which, in turn,
comes from the ``IR safe" combination $k^2\sigma^2 -I(\bk,\br_L)$ in the exponential of the Zel'dovich PS, eq.~\re{PZ}. This observation is all we need to understand what goes wrong when one considers truncated expansions to the full result \re{PZ}, in which the dependence on the IR safe combination is not preserved. Indeed, in this context, RPT-like resummations are obtained by taking out of the $d^3 r_L$ integral the exponentials containing $\sigma^2$, and expanding the one in $I(\bk,\br_L)$ to some finite order, that is, one considers the truncated PS, 
\beqra
&& \!\!\!\!\!\!\!\!\!  \!\!\!\!  \!\!\!\!\!\!\!\!\!  \!\!\!\! \!\!\!\! \!\!\!\! P_Z^N (k;\eta,\etap) \equiv e^{-\frac{k^2 \sigma^2}{2}(e^{\eta}-e^{\etap})^2} \,e^{-k^2\sigma^2 e^{\eta+\etap}}  \int d^3r\; e^{-i\,\bk\cdot\br_L} \sum_{n=1}^N e^{(n-1)(\eta+\etap)}\frac{I(\bk,\br_L)^n}{n!}\,,
\label{PZ-N}
\eeqra
where here, and in what follows, we omit terms proportional to $\delta_D(\bk)$.

The above truncation clearly breaks GI, since none of the $O(q)$ terms of order higher than $N$ in the linear PS coming from the functional derivative of $e^{-k^2\sigma^2 e^{\eta+\etap}}$ can be canceled by the ones coming from the derivatives of the terms in $I(\bk,\br_L)$, which contain at most $N$ powers of $P^0$. Indeed, evaluating the LRF \re{lrf} associated with the truncated PS \re{PZ-N} we obtain
\beqra
&&  \!\!\!\!\!\!\!\!\!  \!\!\!\!  \!\!\!\!\!\!\!\!\!  \!\!\!\! \!\!\!\! \!\!\!\! 
K_{11}^{Z,N} \left( k,q; \eta, \etap \right) =q\, \delta_D(k-q) G_z(k;\eta)G_z(k;\etap)  -  \frac{1}{3} \frac{1}{(2 \pi)^2}  k^2 q \,\left(e^\eta-e^{\etap}\right)^2\, P_Z^N(k;\eta,\etap) \nonumber\\
&&  - \frac{q \, k^2}{6 \pi^2}   \, G_z(k;\eta)G_z(k;\etap) \, \frac{ {\rm e}^{( \eta+\etap)  N} }{N!} 
 \int d^3 r_L \, {\rm e}^{-i \bk \cdot \br_L} \, I \left( \bk,\, \br_L \right)^N  + {\rm O } \left( q^3 \right) \;, 
\label{K11-ZN}
\eeqra
namely an extra ${\rm O } \left( q \right)$ term emerges,  the one at the second line, which does not vanish at equal times and is  of  $N$-th order in $P^0$. This term does not have the form dictated by GI, that is, eq.~\re{lrffullIR}, and it is therefore a spurious $O(q)$ effect induced by the truncation.  

In order to restore the GI  a first strategy could be that of cancelling the spurious IR dependence by adding a proper counterterm. For instance, we can define an ``improved'' truncation, which, for the equal time PS, is given by
\beqra
{\tilde P}_Z^N \left( k; \eta,\, \eta \right) \equiv P_Z^N \left( k; \eta,\, \etap \right) - {\cal K}^{Z,N}_{IR}(k,\eta) \int_0^\Lambda  d q \, P^0 \left( q \right) \;,
\label{Zeld-improved}
\eeqra
where
\beq
{\cal K}^{Z,N}_{IR}(k,\eta) = \lim_{q\to 0} \frac{K_{11}^{Z,N} \left( k,q; \eta, \eta \right)}{q}\,,
\eeq
and $\Lambda$ is a new scale, which is in principle arbitrary, but has to be chosen $\Lambda \alt k$ on physical grounds, otherwise the counterterm would remove also non IR modes which are not constrained by GI.
As we will discuss in the following, the dependence on a new scale is a generic feature of GI invariant resummation schemes. Such dependence vanishes as the truncation order goes to infinity, since the full result must be independent of $\Lambda$.

An alternative strategy is to construct a resummation scheme that is manifestly GI. We start by recalling that the combination $k^2\sigma^2 -I(\bk,\br_L)$  is IR safe, in the sense that the $O(1/q^2)$ terms in the $\int d^3 q$ integrand giving $k^2 \sigma^2$ are canceled by the same terms in the integral giving $I(k,r)$, eq.~\re{I-sigma-def}. A IR-safe quantity can then be obtained by defining 
\begin{equation}
{\bar \sigma}^2 \left( {\bar p} \right) \equiv \frac{1}{6 \pi^2} \int_0^\infty d p \, f \left( \frac{p}{\bar p} \right) P^0 \left( p \right) \;, 
\label{sigma-bar}
\end{equation}
where $f$ is a smooth and integrable function that satisfies $f \left( x \right) \rightarrow 1 + {\rm O } \left( x^2 \right)$ as $x \rightarrow 0$, so that ${\bar \sigma}^2 \left( {\bar p} = \infty \right) = \sigma^2$, but which is otherwise arbitrary. In our explicit computations, the function $f = {\rm e}^{- p^2 / {\bar p}^2}$ will be used. We then replace the expansion \re{PZ-N} by 
\beqra
&&\!\!\!\!\!\!\!\!\! \!\!\!\!\!\!\!\!\!\!\!\!\!\!\!\!\!\! \!\!\!\!\!\!\!\!\!
 {\bar P}_Z^N (k;\eta,\etap)\nonumber\\
 &&\!\!\!\!\!\!\!\!\!  \!\!\!\!\!\!\!\!\! \!\!\!\!\!\!  \!\!\!\!\!\!\!\!\!\equiv e^{-\frac{k^2 \sigma^2}{2}(e^{\eta}-e^{\etap})^2} \,e^{-k^2 \left( \sigma^2 - {\bar \sigma} ^2 \left( {\bar p} \right) \right) e^{\eta+\etap}}  \int d^3r\; e^{-i\,\bk\cdot\br_L} \sum_{n=1}^N e^{(n-1)(\eta+\etap)}\frac{\left[ I \left( \bk, r \right) - k^2 {\bar \sigma}^2 \left( {\bar p} \right) \right]^n }{n!}\,. \nonumber\\ 
\label{PZ-bar-N}
\eeqra

Both truncations \re{PZ-N} and \re{PZ-bar-N} give the  Zel'dovich PS \re{PZ} as $N \rightarrow \infty$. However, the new truncation is GI at all orders. Indeed, evaluating the  equal-time LRF \re{lrf} associated with \re{PZ-bar-N} we obtain, for $q\ll k$,
\beqra
&& \!\!\!\!\!\!\!\!\!  \!\!\!\!\!\!\!\!\!  \!\!\!\!\!\!\!\!\!   \!\!\!\!\!\!\!\!\! 
{\bar K}_{11}^{Z,N} \left( k,q; \eta, \eta \right) =  - \left[ 1 - f \left( \frac{q}{\bar p} \right) \right] \frac{q \, k^2}{6 \pi^2}    {\rm e}^{- k^2 \, \sigma^2 \,  {\rm e}^{2 \eta}} \, \frac{ {\rm e}^{2 \eta  N} }{N!}  \int d^3 r_L \, {\rm e}^{-i \bk \cdot \br_L} \, I \left( \bk,\, \br_L \right)^N  + {\rm O } \left( q^3 \right) \nonumber\\ 
&&   \!\!\!\!\!\!\!\!\! \quad\quad = {\rm O } \left( \frac{q^3 k^2}{{\bar p}^2} \right) + {\rm O } \left( q^3 \right)  \;,
\label{barK11-ZN}
\eeqra
where the last equality holds for any non-vanishing value of $\bar p$.

As we see, no spurious $O(q)$ terms arises, but only $O(q^3)$ ones, which are not protected by GI. The $\bar p$ dependence is of $O((P^0)^N)$, that is $N-1$ loop order, and it is a consequence of the residual arbitrariness in the choice of the resummation scheme once the GI constraints have been satisfied. 

The case $\bar p \rightarrow \infty$ corresponds to SPT, which can then be seen as one special member of a family of GI  expansions. The other special case is the  RPT-like resummation, obtained for $\bar p=0$, which is the only one not GI. It corresponds to the only pathological value for $\bar p$, since the $q\to 0$ and the $\bar p \to 0$ limits do not commute. In more physical terms, eq.~\re{PZ-bar-N} resums the effect of velocity modes for $\bar{p} \alt q < \infty$. Modes with $0\le q \ll k$  have no physical effects on equal-time correlators. This naturally leads to set $\bar p = O(k)$.

\begin{figure}
\centering{ 
\includegraphics[width=0.45\textwidth,clip]{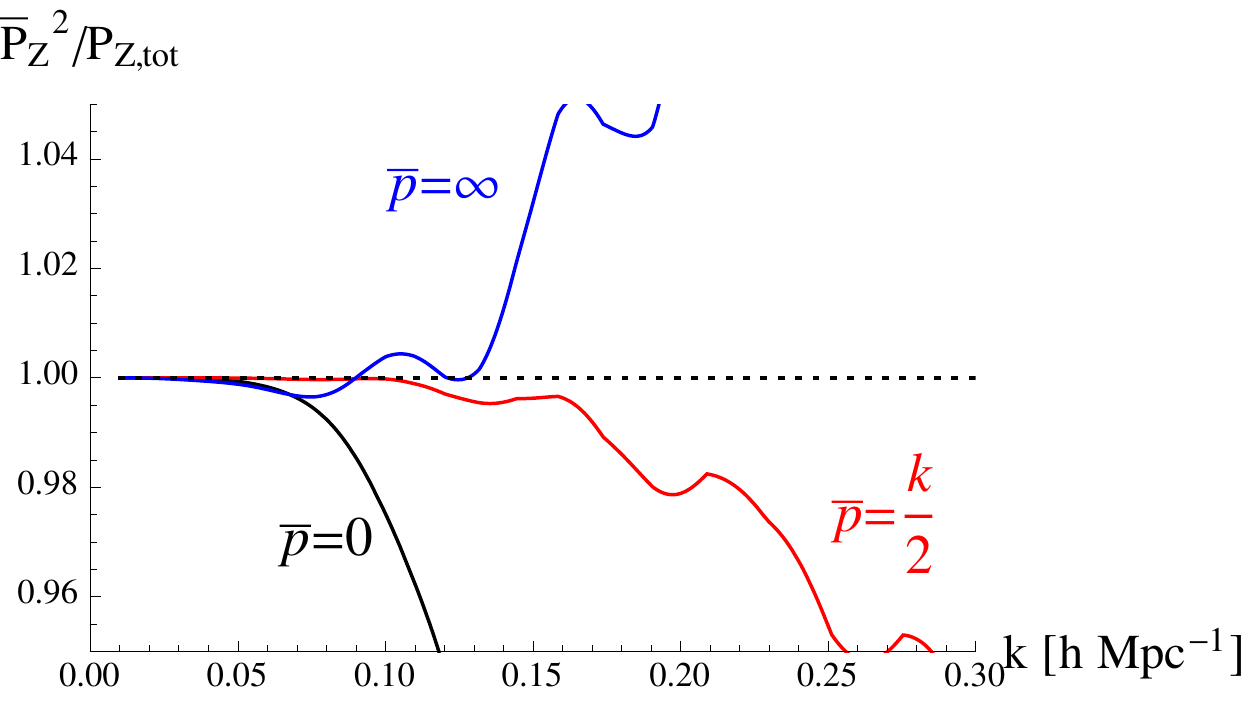}
\includegraphics[width=0.45\textwidth,clip]{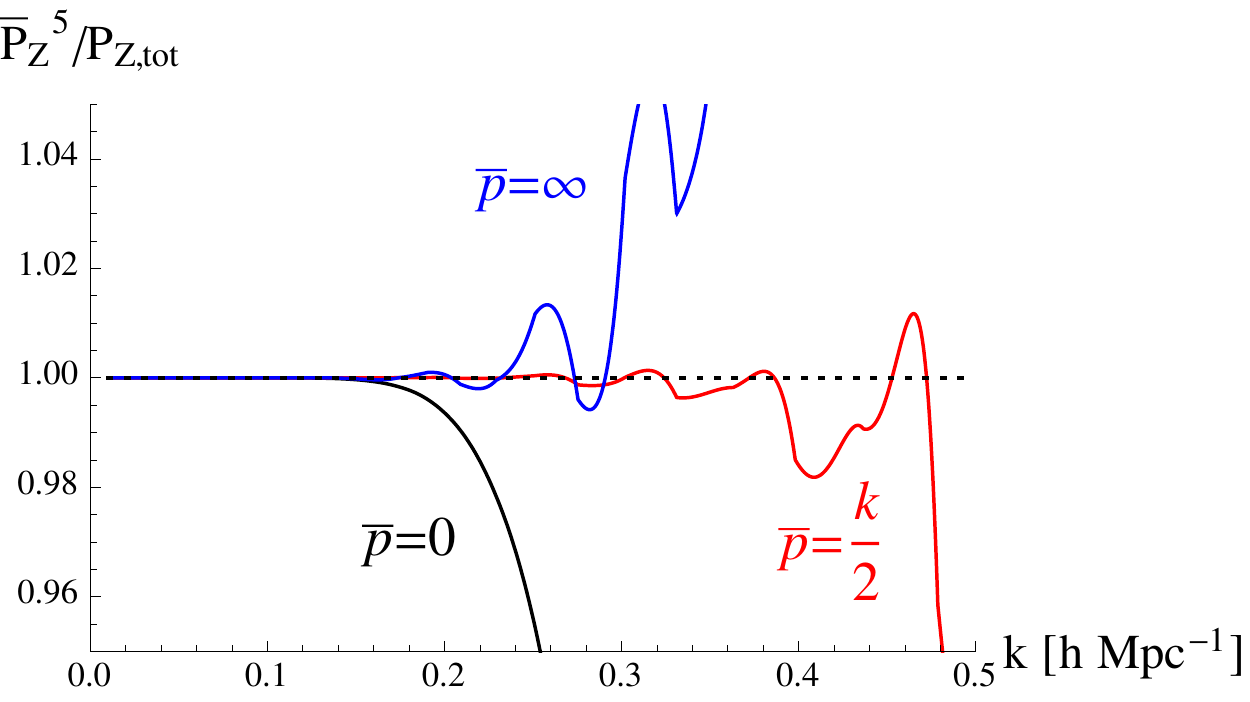}
}
\caption{Ratio between various truncations \re{PZ-bar-N} and the exact density-density equal time power spectrum \re{PZ}. The truncations on the left panel have $N=2$ (corresponding to a one loop diagram in the SPT case), while those in the right panel have $N=5$. We note that the choice ${\bar p} = \frac{k}{2}$ approximates the exact power spectrum better than SPT (${\bar p} = \infty$) and than the non-GI invariant resummation \re{PZ-N} (${\bar p} = 0$). 
}
\label{zeld-compare}
\end{figure}

In Fig.~\ref{zeld-compare} we show the ratio between truncations \re{PZ-bar-N} and the exact density-density equal time power spectrum \re{PZ}. 
The comparison between the two panels shows that, not surprisingly, higher order truncations approximate the exact PS better than the lower order ones. For each given order, we note that the choice  ${\bar p} = \frac{k}{2}$ offers a better approximation than SPT (${\bar p} = \infty$) and than the non-GI invariant resummation \re{PZ-N} (${\bar p} = 0$). 

Before closing this section, we consider the time derivative of \re{PZ-bar-N} with respect to $\eta$, as it will be of use in the following. We get
\beqra
&&\!\!\!\!\!\!\!\!\!\!\!\!\!\!\!\!\!\!\!\!\!\!  \!\!\!\!\!\!\!\!\!\!\! \frac{\partial  {\bar P}_Z^N (k;\eta,\etap) }{\partial \eta} = -k^2 \sigma^2 e^\eta\left(e^\eta-e^{\etap}\right) {\bar P}_Z^N (k;\eta,\etap)-k^2 \left( \sigma^2 - {\bar \sigma} ^2  \left( {\bar p} \right)   \right) e^{\eta+\etap} {\bar P}_Z^N (k;\eta,\etap)\nonumber\\
&&+e^{-\frac{k^2 \sigma^2}{2}(e^{\eta}-e^{\etap})^2} \,e^{-k^2 \left( \sigma^2 - {\bar \sigma} ^2  \left( {\bar p} \right)   \right) e^{\eta+\etap}} \nonumber\\
&&\qquad\qquad \times  \int d^3r\; e^{-i\,\bk\cdot\br_L} \sum_{n=1}^N e^{(n-1)(\eta+\etap)}(n-1) \frac{\left[ I \left( \bk, r\right) - k^2 {\bar \sigma} ^2\left( {\bar p} \right) \right]^n }{n!}\,,\nonumber\\
\eeqra
then, we isolate the 1-loop order contributions from the second and third terms, to obtain 
\beqra
&&\!\!\!\!\!\!\!\!\!\!\!\!\!\!\!\!\!\!\!\!\!\!  \!\!\!\!\!\!\!\!\!\!\! \frac{\partial  {\bar P}_Z^N (k;\eta,\etap) }{\partial \eta} = -k^2 \sigma^2 e^\eta\left(e^\eta-e^{\etap}\right) \left( {\bar P}_Z^N (k;\eta,\etap)- P^0 (k) \right) \nonumber\\
&&\!\!\!\!\!\!\!\!\!\!\!\!\!\!\!\!\!\!\!\!\!\!  \!\!\!\!\!\!\!\!\!\!\!-k^2  \sigma^2 e^{2\eta} P^0 (k) +\frac{e^{\eta+\etap}}{2} \int d^3r\; e^{-i\,\bk\cdot\br_L} I \left( \bk, r \right) ^2 \nonumber\\
&&\!\!\!\!\!\!\!\!\!\!\!\!\!\!\!\!\!\!\!\!\!\!  \!\!\!\!\!\!\!\!\!\!\!-k^2\left(\sigma^2 - {\bar \sigma} ^2\left( {\bar p} \right) \right) e^{\eta+\etap} \left( {\bar P}_Z^N (k;\eta,\etap)  -P^0(k) \right)\nonumber\\
&&\!\!\!\!\!\!\!\!\!\!\!\!\!\!\!\!\!\!\!\!\!\!  \!\!\!\!\!\!\!\!\!\!\! +\frac{e^{\eta+\etap}}{2} \int d^3r\; e^{-i\,\bk\cdot\br_L}  \left( I \left( \bk, r \right)  - k^2 {\bar \sigma} ^2\left( {\bar p} \right) \right)^2\left(e^{-\frac{k^2 \sigma^2}{2}(e^{\eta}-e^{\etap})^2} \,e^{-k^2 \left(\sigma^2 - {\bar \sigma} ^2 \left( {\bar p} \right) \right) e^{\eta+\etap}} -1\right)\nonumber\\
&&\!\!\!\!\!\!\!\!\!\!\!\!\!\!\!\!\!\!\!\!\!\!  \!\!\!\!\!\!\!\!\!\!\! +e^{-\frac{k^2 \sigma^2}{2}(e^{\eta}-e^{\etap})^2} \,e^{-k^2 \left( \sigma^2 - {\bar \sigma} ^2  \left( {\bar p} \right)   \right) e^{\eta+\etap}}\nonumber\\
&&\qquad\qquad\times  \sum_{n=3}^N \int d^3r\; e^{-i\,\bk\cdot\br_L}  e^{(n-1)(\eta+\etap)}(n-1) \frac{\left[ I \left( \bk, r\right) - k^2 {\bar \sigma} ^2\left( {\bar p} \right) \right]^n }{n!}\,,  \nonumber\\
&&
\label{Zeld_res}
\eeqra
where we can check that each line besides the first one (which vanishes as $\etap=\eta$) contains GI-safe combinations of $k^2 \sigma^2$, $k^2 \bar \sigma^2(\bar p)$, and $I(\bk,r)$. The second line is the time-derivative of the Zel'dovich PS in the 1-loop approximation. The third line contains all perturbative orders larger than 1-loop and therefore including it in the equation provides a resummation which coincides with SPT at 1-loop  but contains all the leading terms in the large $k$ limit starting from 2-loop. The remaining three lines include the mode-coupling effects beyond 1-loop and can be systematically added at higher and higher orders. 

In the next section we will consider an evolution equation for the real dynamics analogous to the one above in which the last three lines have been neglected.

\section{Time-Flow equations} 
\label{TRGs}

In this section we consider resummation schemes based on time-flow equations, like the TRG of ref.~\cite{Pietroni08}, eq.~\re{TRGPS}  of the present work, or the one developed in \cite{Anselmi:2010fs,Anselmi:2012cn}, eq.~\re{eq-PS-neq-exact}  of the present work.   To fix ideas, let us start from  eq.~\re{eq-PS-neq-exact}, which, we recall, is exact. In full analogy to what we discussed at the end of the previous section, we want to consider an approximation of this equation which reproduces SPT at small $k$'s and  resums IR effects, relevant for the large-$k$ behavior, in a GI invariant way. We recall that the large-k limit of  the 1-loop $\Sigma_{ab}$ function satisfies $\Sigma^{\mathrm{1-loop}}_{ac} (k;\eta,s)u_c  \to \Sigma^{\mathrm{1-loop,\;eik.}}_{ac}(k;\eta,s)u_c \equiv -k^2 \sigma^2 e^{\eta+s}u_a$. Limiting the SPT order to 1-loop, and taking into account the index structure, we get the analogue of eq.~\re{Zeld_res} with the last three lines neglected, 
\beqra
&& \!\!\!\!\!\!\!\!\!\!\!\!\!\!  \!\!\!\!\!\!\!\! \!\!\!\! 
\partial_\eta P_{ab}(k;\eta,\etap) \simeq \Bigg[ - \Omega_{ac}P_{cb}(k;\eta,\etap) -k^2 \sigma^2 e^\eta \left(e^\eta-e^{\etap}\right)  \left[ P_{ab}(k;\eta,\etap) - P^0(k)u_a u_b \right]  \nonumber\\
&&\!\!\!\! +P^0(k)  u_b \int_{-\infty}^\eta ds\; \Sigma^{\mathrm{1-loop}}_{a c}(k;\eta,s) u_c  + \int_{-\infty}^{\etap}  \,ds \;\Phi^{1-loop}_{ac}(k;\eta,s) g_{bc}(k;\etap,s) \nonumber\\ 
&&\!\!\!\! - k^2 \left[ \sigma^2 -  {\bar \sigma} (\bar p)^2 \right]  \,e^{\eta+\etap}  \left[ P_{ab}(k;\eta,\etap) - P^0(k)u_a u_b \right] \Bigg]
 \;,\nonumber\\
\label{TRG-equalt-GI}
\eeqra
which is GI for any ${\bar p} \neq 0$ (we note that, as in the analogous term in \re{Zeld_res},  the second line is automatically GI, as it comes from the time derivative of the 1-loop SPT expression).

To get the equation for the equal time PS, $P_{ab}(k;\eta,\eta)$, one has to add to the RHS of  \re{TRG-equalt-GI} the quantity which is obtained from the RHS itself by simultaneously swapping the index $a$ with $b$ and the time $\eta$ with $\etap$. 

The solution of the equal-time equation is given by
\beq
P_{ab}(k;\eta,\eta) = P^0(k)u_a u_b + \Delta P^{\mathrm{1-loop}}_{ab}(k;\eta,\eta) + \delta P_{ab}(k;\eta,\eta)\,,
\label{Psol}
\eeq
where $\Delta P^{\mathrm{1-loop}}_{ab}(k;\eta,\eta) $ is the 1-loop SPT contribution to the PS, and
\beqra
&&\!\!\!\!\!\!\!\!\!\!\!\! \!\!\!\!\!\!\!\!\!\!\!\! \!\!\!\!\!\!\! \delta P_{ab}(k;\eta,\eta) = -2 k^2 \left(\sigma^2 -  {\bar \sigma} (\bar p)^2  \right)\nonumber\\
&&\times  \int_{-\infty}^\eta ds\; e^{2s} e^{-k^2  \left(\sigma^2 -  {\bar \sigma} (\bar p)^2  \right)\left(e^{2\eta}-e^{2s}\right)} g_{ac}(\eta,s) g_{bd}(\eta,s) \Delta P^{\mathrm{1-loop}}_{cd}(k;s,s)\,. 
\label{deltaP}
\eeqra
This expression contains contributions of 2-loop and all higher orders, as can be realized by expanding the exponentials in the $O(P^0)$ quantity $\left(\sigma^2 -  {\bar \sigma} (\bar p)^2  \right)$. The temporal integrals can be performed analytically. An interesting limit of this expression is obtained by writing the 1-loop contribution as
\beq
\Delta P^{\mathrm{1-loop}}_{ab}(k;s,s) = e^{2 s} \left(\Delta \bar P^{\mathrm{1-loop}}(k) u_a u_b +  R^{\mathrm{1-loop}}_{ab}(k)\right)\,.
\eeq
If we neglect $R^{\mathrm{1-loop}}_{ab}(k)$ we get
\beq
 \!\!\!\!\!\!\!\!\!\!\!\! \!\!\!\!\!\!\! \delta P_{ab}(k;\eta,\eta) \to  - e^{2\eta} \Delta \bar P^{\mathrm{1-loop}}(k)  u_a u_b \left(1 +\frac{\exp\left(- e^{2 \eta} k^2 \left(\sigma^2 -  {\bar \sigma} (\bar p)^2  \right) \right)-1 }{e^{2\eta} k^2 \left(\sigma^2 -  {\bar \sigma} (\bar p)^2  \right)}\right)\,.
\eeq
It contains again contributions of order 2-loop and higher, but, in the late time limit, the resummation results in a 1-loop contribution which exactly cancels that in \re{Psol}, leaving $P_{ab}(k;\eta,\eta)  \to P^0(k) u_a u_b$. This is to be expected, as $ R^{\mathrm{1-loop}}_{ab}(k)=0$ in the  {\it eikonal} limit, in which only IR-velocity modes are summed \cite{Bernardeau:2012aq, Peloso:2013zw}, and in this limit the equal time nonlinear PS must equal the linear one since, by GI, these modes can have no effect on equal time correlators at shorter scales. This check confirms that eq.~\re{TRG-equalt-GI} does not introduce spurious IR effects. On the other hand, we know that the explicit 1-loop calculation gives a non-vanishing and sizeable $R^{\mathrm{1-loop}}_{ab}(k)$ and, therefore,  \re{deltaP} gives a nontrivial contribution to the nonlinear PS, eq.~\re{Psol}.
\begin{figure}
\centering{ 
\includegraphics[width=0.45\textwidth,clip]{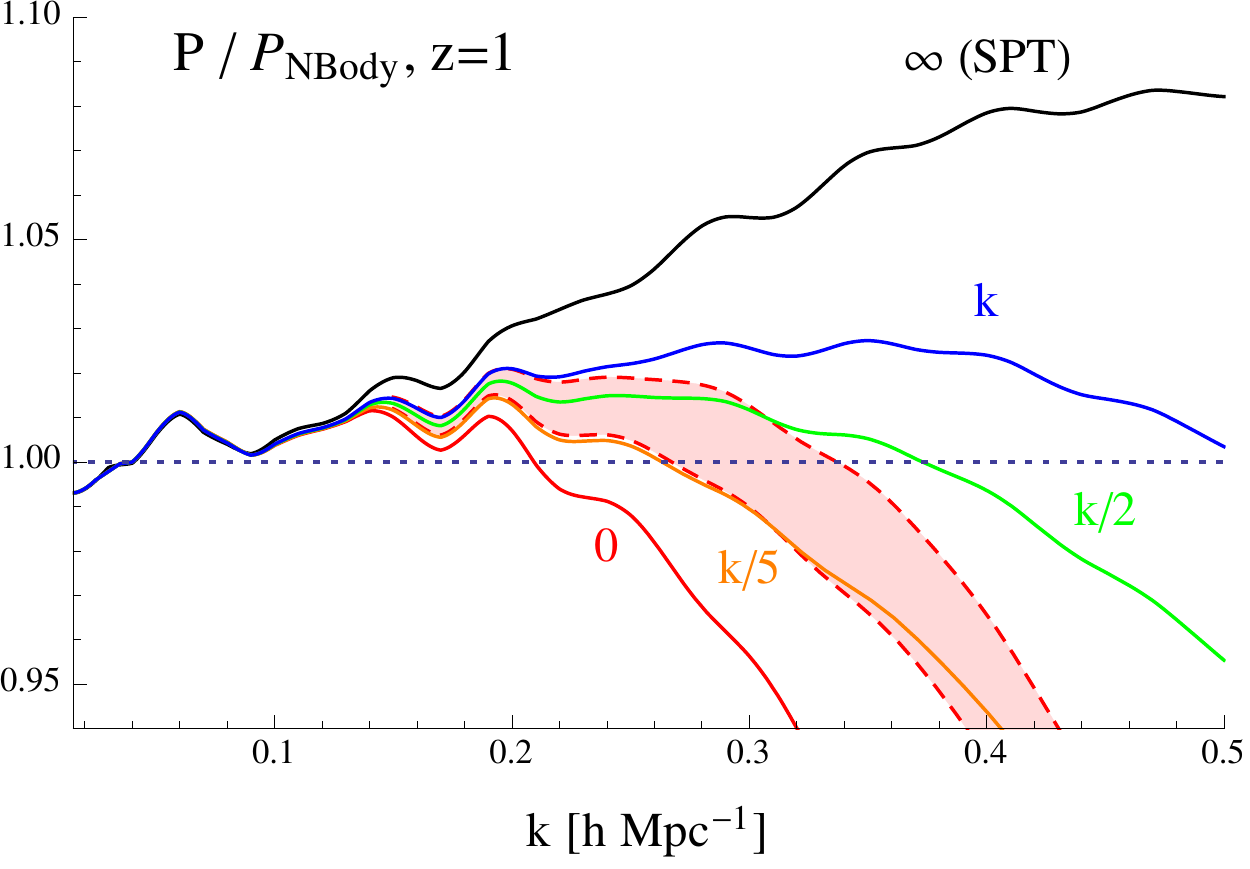}
\includegraphics[width=0.45\textwidth,clip]{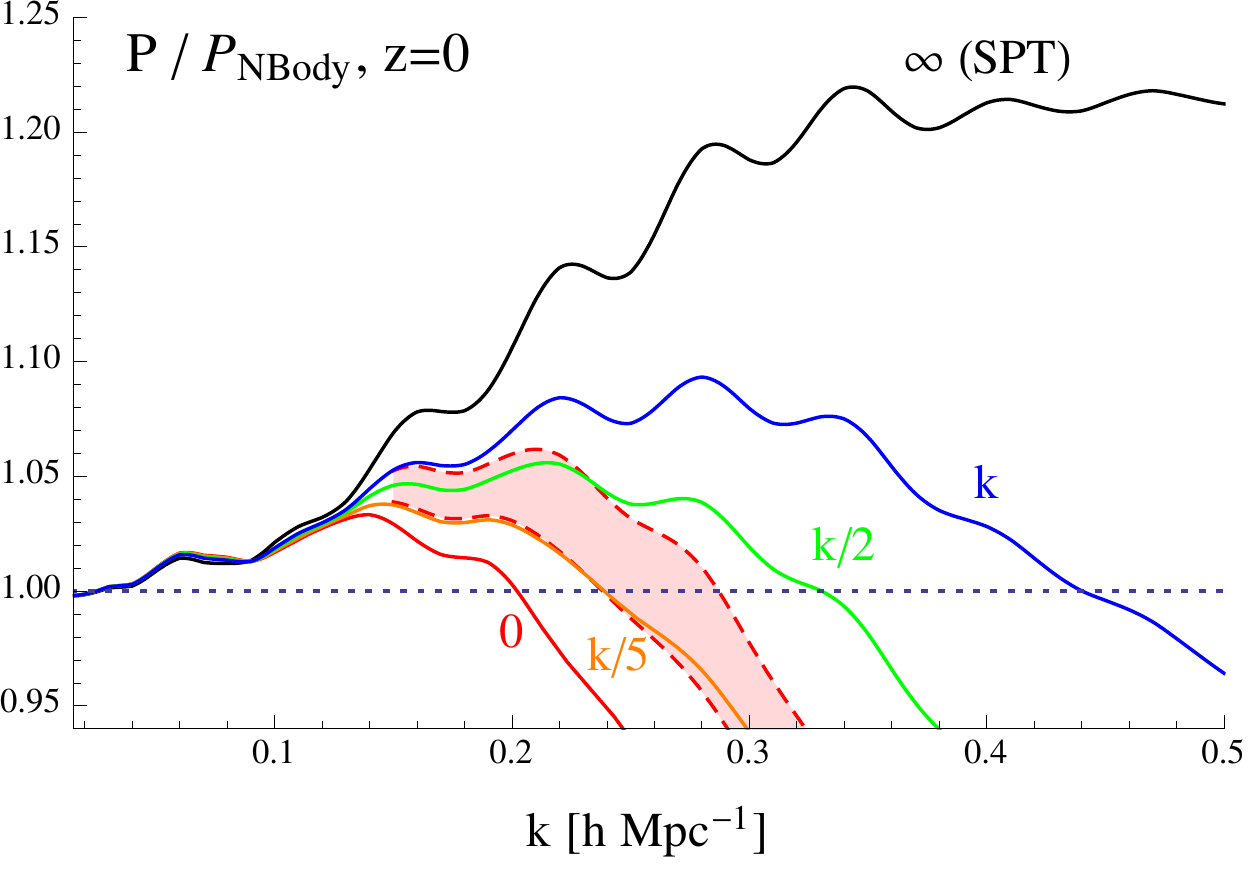}
}
\caption{Ratio between the density-density PS obtained from the TRG equation \re{TRG-equalt-GI} and the one obtained in numerical simulations. 
Different solid lines are obtained with different values of ${\bar p}$, as indicated by the labels. The shaded region is obtained from first solving the TRG equation with ${\bar p} = 0$, and then by ``improving'' the result as in equation \re{Zeld-improved}, with $\Lambda$ varying from $k/5$ to $k$ across the shaded region. We note that the solid line with $k/5$ is hardly visible, as it is inside this shaded region. The left (respectively, right) panel shows the ratio at redshift $z=1$ (respectively, $z=0$). 
}
\label{TRG}
\end{figure}

In Fig.~\ref{TRG} we compare the density-density power spectrum $P_{11}$ obtained with the TRG equation (\ref{TRG-equalt-GI}) against the nonlinear  power spectrum given by the the FrankenEmu \cite{Heitmann:2013bra} N-body based emulator.  The spectra are computed for the ``REF'' cosmology simulation of \cite{Manzotti:2014loa} (see Section 4 of that work), with initial condition at $\eta = \etain$ (corresponding to $z = z_{\rm in} = 99$) generated from the linear CAMB code \cite{Lewis:1999bs}.  The labels on the solid curves refer to the value of ${\bar p}$ used in that evolution (we recall that ${\bar \sigma}^2 \left( {\bar p} \right)$ is chosen as in eq. (\ref{sigma-bar}) with $f = {\rm e}^{-p^2 / {\bar p}^2}$). The choice ${\bar p} =\infty$ corresponds to ${\bar \sigma}^2 \left( \infty \right) = \sigma^2$, so that the last line of  (\ref{TRG-equalt-GI}) vanishes, and our equation reproduces the SPT 1-loop result. We then show results for decreasing values of ${\bar p} \la k$. Finally, the ${\bar p} = 0$ case corresponds to ${\bar \sigma}^2 \left( 0 \right) = 0$, namely with a non GI resummation. 

A way to ``improve'' this non-GI (${\bar \sigma}^2 \left( 0 \right) = 0$) result is to add a counterterm identical to eq. \re{Zeld-improved}. More precisely, we numerically evaluate the LRF \re{lrf} $K_{11} \left( k, q_{\rm min} ; \eta, \eta \right)$ at varying $k$ and at fixed $q_{\rm min} = 10^{-4} \, h \, {\rm Mpc}^{-1}$.~\footnote{To do so, we evaluate the loop integrals for $\Sigma^{\mathrm{1-loop}}$ and for $\Phi^{1-loop}$, as well as the integral \re{I-sigma-def} for $\sigma^2$, for the five different values $q_{\rm min} = \left\{ 0.8 ,\, 0.9 ,\, 1 ,\, 1.1 ,\, 1.2 \right\} \times 10^{-4}  \, h \, {\rm Mpc}^{-1}$ of the lower extremum of integration (the upper extremum of integration is instead fixed to $q_{\rm max} = 10 \,  \, h \, {\rm Mpc}^{-1}$). We then use these quantities in the TRG equation (\ref{TRG-equalt-GI}),  solve for the power spectrum, and use the results to  evaluate the first five coefficients in the Taylor expansion $P_{11}  \left( \bk; q_{\rm min}  \right) = \sum_{k=0}^5 \frac{1}{k!} \, \frac{d^k P_{11}}{d q_{\rm min}^k} \vert_{{q_{\rm min}} =  10^{-4}   \, h \, {\rm Mpc}^{-1}} \times \left( q_{{\rm min}} - 10^{-4}   \, h \, {\rm Mpc}^{-1} \right)^k$; the first derivative gives the LRF. Using $5$ points (rather than for example ony $2$) decreases the numerical noise.} The improved PS has an arbitrary scale $\Lambda$ (see eq.  \re{Zeld-improved}). Varying $\Lambda$ from $k/5$ to $k$ gives an improved power spectrum in the shaded region in the figure. 

As in the Zel'dovich example shown in Fig.~\ref{zeld-compare}, the GI result obtained with ${\bar p} = k/2$ agrees with the numerical simulations much better than the SPT result (${\bar p} = \infty$), and better than the non-GI case  (${\bar p} = 0$). For instance, at $z=1$, the SPT result is within $6 \%$ from the numerical simulation for $k \la 0.32  \, h \, {\rm Mpc}^{-1}$. On the contrary, the TRG result obtained with ${\bar p} = k/2$ is within  $6 \%$ for all the values of $k$ shown in the figure. At $z=0$, the agreement is within $6\%$ for $k \la  0.14  \, h \, {\rm Mpc}^{-1}$ in the SPT case, and for $k \la  0.38  \, h \, {\rm Mpc}^{-1}$ in the TRG-${\bar p} = k/2$ case. 

Another, possibly more powerful, test of  IR resummation methods is to compute the PS at different times, $P(k;\eta,\etap)$. This quantity is not observable, given that our observations are limited to our past light-cone, but it is nevertheless physical, in the sense that it is measurable, for instance, by cross correlating different time-snapshots in a simulation. From the theoretical point of view, considering this quantity gives access to the $O(q)$ term in \re{lrffullIR}, which, unlike the $O(q^3)$ ones, is fixed by GI nonperturbatively. Therefore, we  consider eq.~\re{TRG-equalt-GI}, for $P(k;\eta,\etap)$ with $\etap=\etain\to - \infty$,
\beqra
&& \!\!\!\!\!\!\!\!\!\!\!\!\!\!  \!\!\!\!\!\!\!\!  \!\!\!\!\!\!\!\! 
\partial_\eta P_{ab}(k;\eta,-\infty) 
 \simeq -\Omega_{ac}P_{cb}(k;\eta,-\infty) +P^0(k)  u_b \int_{-\infty}^\eta ds\; \Sigma^{\mathrm{1-loop}}_{a c}(k;\eta,s) u_c \nonumber\\
&&   -k^2 \sigma^2 \,e^{2\eta}  \left( P_{ab}(k;\eta,-\infty) -P^0(k)u_a u_b \right) \,,  \nonumber\\
\label{eq-PS-neq-approx1}
\eeqra
which interpolates between the 1-loop SPT result, and the large-k result for the propagator (as we can see from the relation \re{Pp-PMC}, the quantity $P_{ab}(k;\eta,\etain)$ is equal to  $G_{ac} \left( k; \eta , \etain \right) u_c \, u_b P^0 \left( k \right)$ if there are no primordial nongaussianities (encoded in the $\Phi_{ab}$ function) at $\etain\to -\infty$ ). Notice that, since all the momentum modes affect the nonequal time PS, the $\bar\sigma^2(\bar p)$ dependence drops out of the equation. This physical fact is also the origin for the $O(q)$ dependence of \re{lrffullIR}. 

The solution of this equation is given by
\beq
 \!\!\!\!\!\!\!\!  \!\!\!\!\!\!\!\!   \!\!\!\!\! P_{ab}(k;\eta,-\infty)  = P^0(k) u_a u_b + \Delta G_{ac}^{\mathrm{1-loop}}(k;\eta,-\infty) u_c P^0(k) u_b + \delta P_{ab}(k;\eta,-\infty) \,,
 \label{Gsol}
\eeq
where the first two terms are the 1-loop SPT result (namely the solution of the first line of \re{eq-PS-neq-approx1}), with 
\beq
 \!\!\!\!\!\!\!\!  \!\!\!\!\!\!\!\!   \!\!\!\!\! 
\Delta G_{ac}^{\mathrm{1-loop}}(k;\eta,-\infty) \, u_c \equiv \int_{-\infty}^\eta d s \, g_{ab} \left( \eta-s \right) \int_{-\infty}^s d s' \, 
\Sigma^{\mathrm{1-loop}}_{b c}(k;s,s') \, u_c \;, \nonumber\\
\eeq
and
\beqra
&&\!\!\!\!\!\!\!\!\!\!\!\! \!\!\!\!\!\!\!\!\!\!\!\! \!\!\!\!\!\!\! \!\!\!\! \delta P_{ab}(k;\eta,-\infty ) = - k^2\sigma^2  P^0(k) u_b \int_{-\infty}^\eta ds\; e^{2s} e^{-k^2  \sigma^2 \frac{\left(e^{2\eta}-e^{2s}\right)}{2}} g_{ac}(\eta,s) \Delta G_{cd}^{\mathrm{1-loop}}(k,s,-\infty) u_d \,. \nonumber\\
\label{deltaG}
\eeqra
Since $ \Delta G_{cd}^{\mathrm{1-loop}}(k,s,-\infty) \propto e^{2s}$ also in this case the time integrals can be performed analytically. In the eikonal limit we have $\Delta G_{ac}^{\mathrm{1-loop}}(k;\eta,-\infty) u_c \to - k^2 \sigma^2 \frac{e^{2 \eta}}{2} u_a\,,$ which, inserted in \re{deltaG} gives
\beq
\delta P_{ab}(k;\eta,-\infty ) \to P^0(k) u_a u_b \left(-1 +  k^2 \sigma^2 \frac{e^{2 \eta}}{2} + \exp\left(-  k^2 \sigma^2 \frac{e^{2 \eta}}{2}  \right)  \right)\,,
\eeq
and therefore we recover from \re{Gsol} the eikonal, or Zel'dovich, result, $P_{ab}(k;\eta,-\infty) \to P^0(k) u_a u_b  \exp\left(-  k^2 \sigma^2 \frac{e^{2 \eta}}{2}  \right) $.

\begin{figure}
\centering{ 
\includegraphics[width=0.45\textwidth,clip]{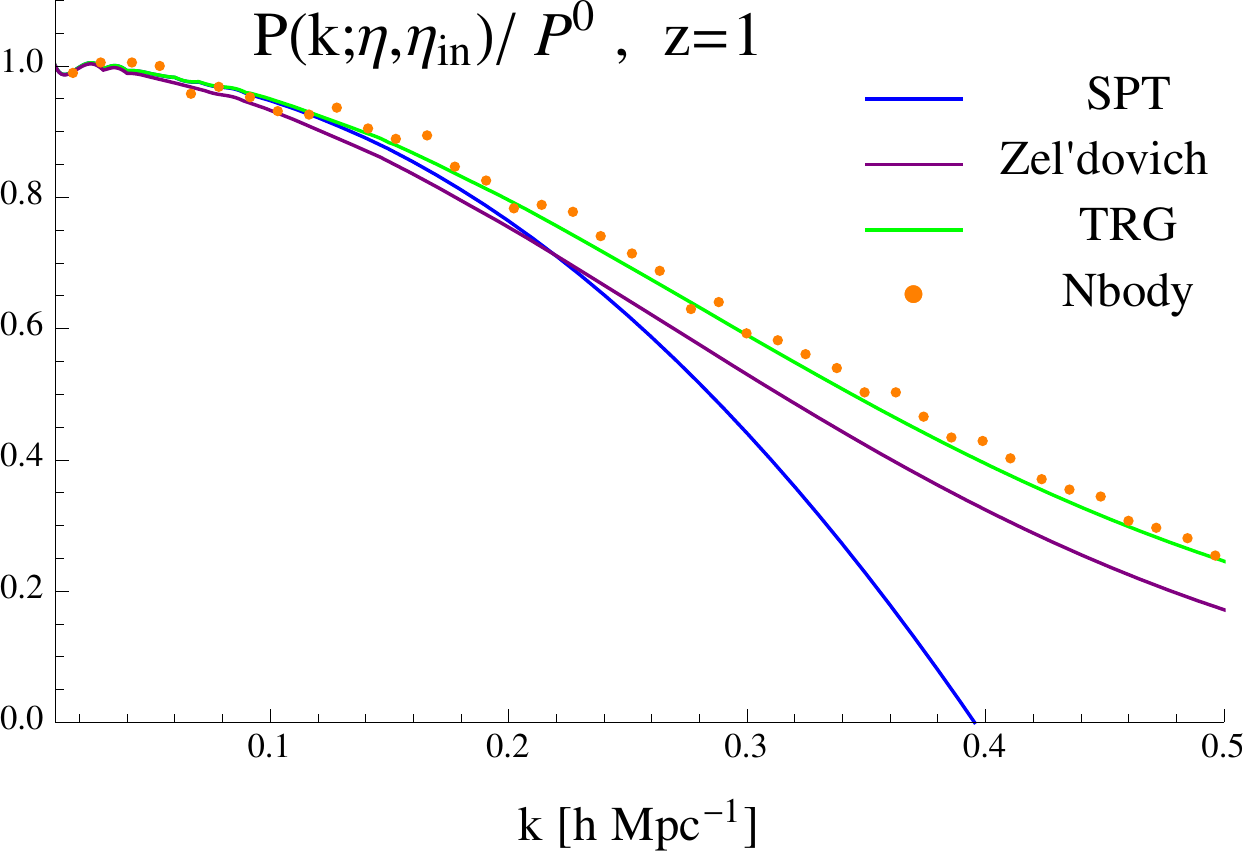}
\includegraphics[width=0.45\textwidth,clip]{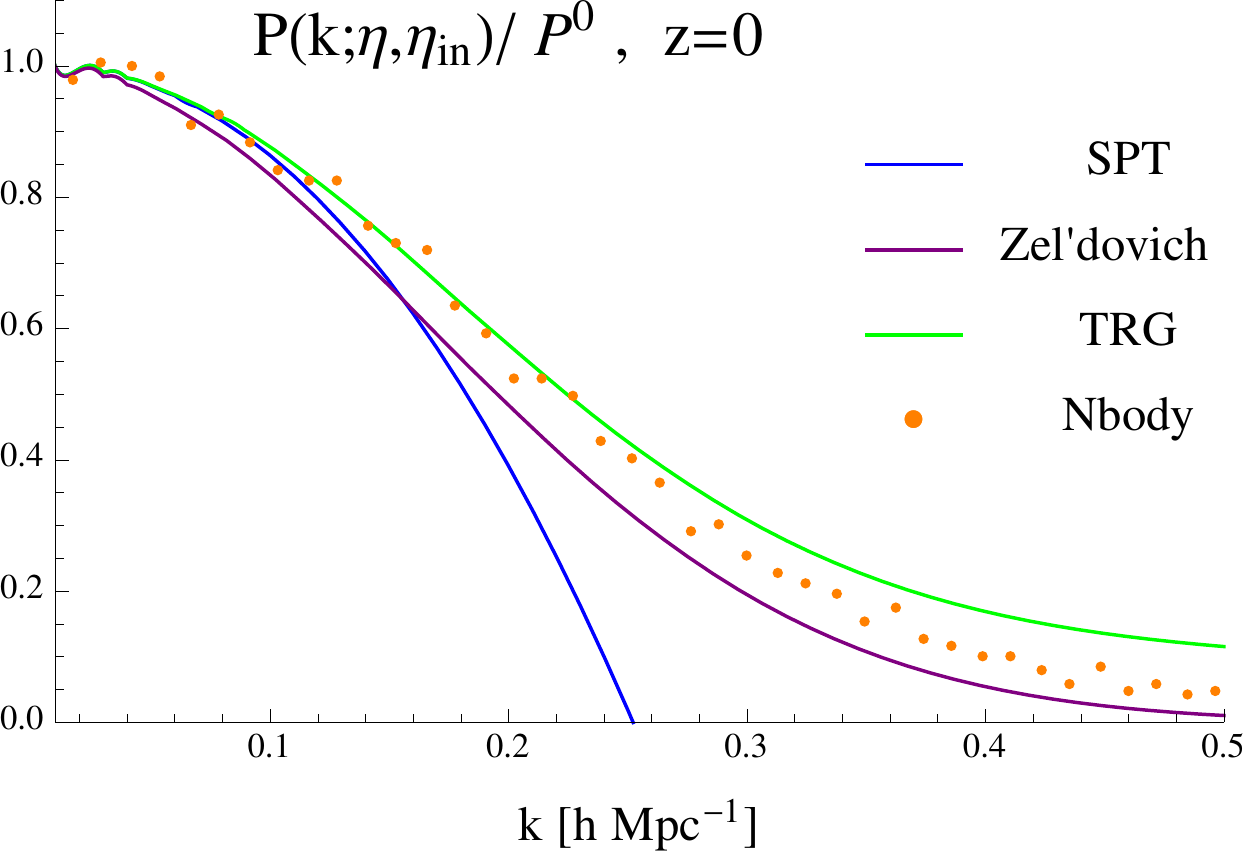}
}
\caption{Comparison between N-body data and analytic approximations of the density-density unequal time power spectrum $P_{11} \left( k; \eta, \etain \right)$, where $\eta$ corresponds to $z=1$ and $z=0$ in the left and right panel, respectively, and where $\etain$ is taken at $z=99$ (which can be taken to be $-\infty$ in the analytic computations). The power spectrum shown is divided by the linear power spectrum. The ratio  corresponds to the sum of the two components $G_{11} \left( k; \eta , -\infty \right) + G_{12} \left( k; \eta , -\infty \right) $ of the exact propagator $G$. The N-body data, and the analytics approximations, are described in the  text. 
}
\label{fig:propagator}
\end{figure}

In Fig. \ref{fig:propagator} we show the density-density non-equal time PS $P_{11} \left( k; \eta , -\infty \right)$, divided by the linear power spectrum. As we already remarked, this corresponds to the sum of the two propagator components  $G_{11} \left( k; \eta , -\infty \right) + G_{12} \left( k; \eta , -\infty \right) $. The N-body data are obtained from the  ``REF'' cosmology simulation of \cite{Manzotti:2014loa}. These data are compared against the 1-loop SPT approximation (namely, the first two terms in (\ref{Gsol})), the Zel'dovich approximation $ \exp\left(-  k^2 \sigma^2 \frac{e^{2 \eta}}{2}  \right) $, and our TRG solution (namely, the full expression in (\ref{Gsol})). The left panel of the figure is obtained at redshift $z=1$. We note that in this case  the TRG solution describes the N-body data better than the SPT and the Zel'dovich approximations, and  with a good accuracy at all momenta shown in the figure. The accuracy worsens at $z=0$, as can be seen from the right  panel. We note that, in this case, the TRG solution accurately follows the data up to $k \la 0.25 h {\rm Mpc}^{-1}$, where it performs better than the SPT and the Zel'dovich approximation. 
We also note that the TRG solution approaches the Zel'dovich  one only at momenta greater than the ones shown in the figure.

\section{Conclusions}
\label{conclusions}

In this work we studied the invariance under uniform, but time-dependent, boosts (GI, in short) of computational schemes for the large scale structures in the universe. Many resummation schemes that aim for an improvment over standard perturbation theory break GI, and are therefore affected by a spurious IR-UV connection. For instance, the computation of the PS at a scale $k$ is affected by long velocity modes at scales  $\bar p \alt k$, through terms depending on $k^2 \bar \sigma(\bar p)^2$, with $\sigma(\bar p)^2$  the velocity dispersion of the long modes.

The effect of mode-mode coupling can be studied through response functions. In this work we studied the linear response of the nonlinear PS, $P (k;\eta,\etap)$, evaluated at a scale $k$, to changes of the initial conditions at a much larger scale  $q^{-1} \gg k^{-1}$, namely $K \left( k, q;\eta,\etap \right) \propto q \, \frac{\delta P \left( k;\eta,\etap \right)}{\delta P^0 \left( q \right)}$. By using the consistency relations of  \cite{Peloso:2013zw}, we found that GI demands that~\footnote{This relation is for the density-density power spectrum; see eq. (\ref{lrffullIR}) for correlators that involve also the velocity gradient.} 
\begin{equation}
\!\!\!\!\!\!\!\! \!\!\!\!\!\!\!\! \!\!\!\!\!\!\!\! 
K (k,q;\eta,\etap) =  -  \frac{1}{3} \frac{1}{(2 \pi)^2}  k^2 q \,\left(e^\eta-e^{\etap}\right)^2\, P (k;\eta,\etap) + O(q^3) \;\; , \;\; q \ll k \,. 
\label{LRF-IR-conclusions}
\end{equation} 
The $O(q)$ coefficient in front of the fully nonlinear PS is protected by GI, and is therefore not renormalized at any order in SPT and even beyond that, when shell-crossing or virialization occurs. At equal times, $\eta=\etap$, it vanishes, and we are left with $O(q^3)$ terms whose coefficients are left undetermined by GI. Therefore, the strictest  tests of GI for a given approximation scheme are obtained by considering non-equal time correlation functions. We have seen that the Zel'dovich PS passes the test fully, while SPT does it only up to the perturbative order of the calculation, which is the case of its poor performance on the propagator, see Fig.~\ref{fig:propagator}, or in accounting for the widening of the BAO peak. 

Non GI approximation schemes are characterized by having extra $O(q)$ terms, which can even be non-vanishing in the equal time limit. The coefficients of these terms can be used to tailor a counterterm  which, when added to the original PS, gives an improved one with the correct behavior for the linear response function, and a better convergence to the full result, see for instance eq.~\re{Zeld-improved}.

The arbitrariness of the coefficient of the $O(q^3)$ term reflects the fact that there are many possible GI approximation schemes: Eulerian PT, Lagrangian PT, eRPT \cite{Anselmi:2012cn, Peloso:2013zw}, Time-Sliced PT \cite{Blas:2016sfa}, IR resummed effective field theory \cite{Senatore:2014via}, and so on, all of which satisfy eq.~\re{LRF-IR-conclusions}, and differ precisely on that coefficient. This, effectively, translates in a dependence of the result - at a finite order in the given approximation scheme - on some new scale. This dependence vanishes by increasing the truncation order to infinity, in which limit all the different schemes should converge to the same result (at least as far as the IR effects are concerned). In this sense, the $\bar p$ scale introduced in this paper  (as well as the detailed momentum dependence of the cutoff function in eq.~\re{sigma-bar}) ``mimics'' the scheme dependence of finite order results in different GI approximation schemes. We have introduced a family of approximation schemes parameterized by $\bar p$,  from SPT (for ${\bar p} = \infty$, so that there is no resummation) to the physical range ${\bar p} \la k$, which  avoids the spurious effects from modes of scales with  $q \ll k$. We have also seen that RPT-like resummations correspond to the special value $\bar p = 0$, which singles out the only non GI scheme in the family parameterized by $\bar p$.

After discussing the GI issue in the Zel'dovich approximation, we moved to the exact dynamics (still in the single stream approximation). We focused on the Time-flow equations, like the TRG approach \cite{Pietroni08} or the equations discussed in \cite{Anselmi:2010fs,Anselmi:2012cn}, clarifying the relation between the two (see eqs. \re{TRGPS} and \re{eq-PS-neq-exact}). We derived GI TRG equations which incorporate IR resummations and match to finite order SPT at small $k$'s.    In Figure \ref{TRG} we show that, even limiting the matching to SPT  to 1-loop order, setting the cut-off scale ${\bar p} \la k$ approximates the exact power spectrum (provided in this case by N-body simulation) significantly better than the original TRG and the plain one-loop SPT. The Time-Flow equations that we have introduced can be solved analytically  in the $\Lambda$CDM case, as we have done. Moreover, they provide a convenient approach for cosmologies that are characterized by a scale-dependent growth factor, as for instance in the case of massive neutrinos \cite{LMPR09}, in which these equations can be easily solved numerically. 

The main goal of this work was to discuss on general grounds how to incorporate IR effects in a resummation scheme and to show how maintaining  GI can improve their accuracy. We explicitly verified this up to four loop level in the Zel'dovich approximation, and up to one loop level in the exact case (in single stream approximation). Once the IR sector is fixed, our TRG equations can be further improved by considering mode-mode coupling at higher loops, and by including UV effects, as for instance in the coarse-grained approach of \cite{Pietroni:2011iz, Manzotti:2014loa}. We plan to come back to this in a future publication.

\section*{Acknowledgments}
M. Pietroni acknowledges support from the European Union Horizon 2020 research and innovation programme under the Marie Sklodowska-Curie grant agreements Invisible- sPlus RISE No. 690575, Elusives ITN No. 674896 and Invisibles ITN No. 289442.
The work of M. Peloso was supported in part by DOE grant de-sc0011842 at the University of Minnesota.

\section*{References}
\bibliographystyle{JHEP}
\bibliography{/Users/massimo/Dropbox/Mnu/PostTRG/draft/mybib.bib}

\providecommand{\href}[2]{#2}\begingroup\raggedright\begin{thebibliography}{10}

\bibitem{PT}
F.~Bernardeau, S.~Colombi, E.~Gaztanaga and R.~Scoccimarro, {\it {Large-scale
  structure of the universe and cosmological perturbation theory}},  {\em Phys.
  Rept.} {\bf 367} (2002) 1--248
  [\href{http://arXiv.org/abs/astro-ph/0112551}{{\tt astro-ph/0112551}}].

\bibitem{JK06}
D.~Jeong and E.~Komatsu, {\it {Perturbation Theory Reloaded: Analytical
  Calculation of Non-linearity in Baryonic Oscillations in the Real Space
  Matter Power Spectrum}},  {\em Astrophys. J.} {\bf 651} (2006) 619
  [\href{http://arXiv.org/abs/astro-ph/0604075}{{\tt astro-ph/0604075}}].

\bibitem{Blas:2013aba}
D.~Blas, M.~Garny and T.~Konstandin, {\it {Cosmological perturbation theory at
  three-loop order}},  {\em JCAP} {\bf 1401} (2014), no.~01 010
  [\href{http://arXiv.org/abs/1309.3308}{{\tt 1309.3308}}].

\bibitem{Little:1991py}
B.~Little, D.~H. Weinberg and C.~Park, {\it {Primordial fluctuations and
  nonlinear structure}},  {\em Mon. Not. Roy. Astron. Soc.} {\bf 253} (1991)
  295--306.

\bibitem{2014arXiv1411.2970N}
T.~{Nishimichi}, F.~{Bernardeau} and A.~{Taruya}, {\it {Response function of
  the large-scale structure of the universe to the small scale
  inhomogeneities}},  {\em ArXiv e-prints} (Nov., 2014)
  [\href{http://arXiv.org/abs/1411.2970}{{\tt 1411.2970}}].

\bibitem{Pietroni:2011iz}
M.~Pietroni, G.~Mangano, N.~Saviano and M.~Viel, {\it {Coarse-Grained
  Cosmological Perturbation Theory}},  {\em JCAP} {\bf 1201} (2012) 019
  [\href{http://arXiv.org/abs/1108.5203}{{\tt 1108.5203}}].

\bibitem{Manzotti:2014loa}
A.~Manzotti, M.~Peloso, M.~Pietroni, M.~Viel and F.~Villaescusa-Navarro, {\it
  {A coarse grained perturbation theory for the Large Scale Structure, with
  cosmology and time independence in the UV}},  {\em JCAP} {\bf 1409} (2014),
  no.~09 047 [\href{http://arXiv.org/abs/1407.1342}{{\tt 1407.1342}}].

\bibitem{Baumann:2010tm}
D.~Baumann, A.~Nicolis, L.~Senatore and M.~Zaldarriaga, {\it {Cosmological
  Non-Linearities as an Effective Fluid}},  {\em JCAP} {\bf 1207} (2012) 051
  [\href{http://arXiv.org/abs/1004.2488}{{\tt 1004.2488}}].

\bibitem{Carrasco:2012cv}
J.~J.~M. Carrasco, M.~P. Hertzberg and L.~Senatore, {\it {The Effective Field
  Theory of Cosmological Large Scale Structures}},  {\em JHEP} {\bf 1209}
  (2012) 082 [\href{http://arXiv.org/abs/1206.2926}{{\tt 1206.2926}}].

\bibitem{Blas:2015tla}
D.~Blas, S.~Floerchinger, M.~Garny, N.~Tetradis and U.~A. Wiedemann, {\it
  {Large scale structure from viscous dark matter}},  {\em JCAP} {\bf 1511}
  (2015) 049 [\href{http://arXiv.org/abs/1507.06665}{{\tt 1507.06665}}].

\bibitem{Floerchinger:2016hja}
S.~Floerchinger, M.~Garny, N.~Tetradis and U.~A. Wiedemann, {\it
  {Renormalization-group flow of the effective action of cosmological
  large-scale structures}},  \href{http://arXiv.org/abs/1607.03453}{{\tt
  1607.03453}}.

\bibitem{RPTb}
M.~Crocce and R.~Scoccimarro, {\it {Memory of Initial Conditions in
  Gravitational Clustering}},  {\em Phys. Rev.} {\bf D73} (2006) 063520
  [\href{http://arXiv.org/abs/astro-ph/0509419}{{\tt astro-ph/0509419}}].

\bibitem{RPTa}
M.~Crocce and R.~Scoccimarro, {\it {Renormalized Cosmological Perturbation
  Theory}},  {\em Phys. Rev.} {\bf D73} (2006) 063519
  [\href{http://arXiv.org/abs/astro-ph/0509418}{{\tt astro-ph/0509418}}].

\bibitem{Bernardeau:2008fa}
F.~Bernardeau, M.~Crocce and R.~Scoccimarro, {\it {Multi-Point Propagators in
  Cosmological Gravitational Instability}},  {\em Phys.Rev.} {\bf D78} (2008)
  103521 [\href{http://arXiv.org/abs/0806.2334}{{\tt 0806.2334}}].

\bibitem{Anselmi:2010fs}
S.~Anselmi, S.~Matarrese and M.~Pietroni, {\it {Next-to-leading resummations in
  cosmological perturbation theory}},  {\em JCAP} {\bf 1106} (2011) 015
  [\href{http://arXiv.org/abs/1011.4477}{{\tt 1011.4477}}].

\bibitem{Anselmi:2012cn}
S.~Anselmi and M.~Pietroni, {\it {Nonlinear Power Spectrum from Resummed
  Perturbation Theory: a Leap Beyond the BAO Scale}},  {\em JCAP} {\bf 1212}
  (2012) 013 [\href{http://arXiv.org/abs/1205.2235}{{\tt 1205.2235}}].

\bibitem{Matsubara07}
T.~Matsubara, {\it {Resumming Cosmological Perturbations via the Lagrangian
  Picture: One-loop Results in Real Space and in Redshift Space}},  {\em Phys.
  Rev.} {\bf D77} (2008) 063530 [\href{http://arXiv.org/abs/0711.2521}{{\tt
  0711.2521}}].

\bibitem{Turb_Pope}
S.~B. Pope, {\em Turbulent Flows}.
\newblock Cambridge University Press, August, 2000.

\bibitem{Scoccimarro:1995if}
R.~Scoccimarro and J.~Frieman, {\it {Loop corrections in nonlinear cosmological
  perturbation theory}},  {\em Astrophys.J.Suppl.} {\bf 105} (1996) 37
  [\href{http://arXiv.org/abs/astro-ph/9509047}{{\tt astro-ph/9509047}}].

\bibitem{Peloso:2013zw}
M.~Peloso and M.~Pietroni, {\it {Galilean invariance and the consistency
  relation for the nonlinear squeezed bispectrum of large scale structure}},
  {\em JCAP} {\bf 1305} (2013) 031 [\href{http://arXiv.org/abs/1302.0223}{{\tt
  1302.0223}}].

\bibitem{Pietroni08}
M.~Pietroni, {\it {Flowing with Time: a New Approach to Nonlinear Cosmological
  Perturbations}},  {\em JCAP} {\bf 0810} (2008) 036
  [\href{http://arXiv.org/abs/0806.0971}{{\tt 0806.0971}}].

\bibitem{Senatore:2014via}
L.~Senatore and M.~Zaldarriaga, {\it {The IR-resummed Effective Field Theory of
  Large Scale Structures}},  {\em JCAP} {\bf 1502} (2015), no.~02 013
  [\href{http://arXiv.org/abs/1404.5954}{{\tt 1404.5954}}].

\bibitem{Baldauf:2015xfa}
T.~Baldauf, M.~Mirbabayi, M.~Simonovi{\'c} and M.~Zaldarriaga, {\it
  {Equivalence Principle and the Baryon Acoustic Peak}},  {\em Phys. Rev.} {\bf
  D92} (2015), no.~4 043514 [\href{http://arXiv.org/abs/1504.04366}{{\tt
  1504.04366}}].

\bibitem{Blas:2016sfa}
D.~Blas, M.~Garny, M.~M. Ivanov and S.~Sibiryakov, {\it {Time-Sliced
  Perturbation Theory II: Baryon Acoustic Oscillations and Infrared
  Resummation}},  {\em JCAP} {\bf 1607} (2016), no.~07 028
  [\href{http://arXiv.org/abs/1605.02149}{{\tt 1605.02149}}].

\bibitem{MP07b}
S.~Matarrese and M.~Pietroni, {\it {Resumming Cosmic Perturbations}},  {\em
  JCAP} {\bf 0706} (2007) 026
  [\href{http://arXiv.org/abs/astro-ph/0703563}{{\tt astro-ph/0703563}}].

\bibitem{2013MNRAS.428.3173J}
G.~{J{\"u}rgens} and M.~{Bartelmann}, {\it {Perturbation theory trispectrum in
  the time renormalization approach}},  {\em MNRAS} {\bf 428} (Feb., 2013)
  3173--3182.

\bibitem{Peloso:2013spa}
M.~Peloso and M.~Pietroni, {\it {Ward identities and consistency relations for
  the large scale structure with multiple species}},  {\em JCAP} {\bf 1404}
  (2014) 011 [\href{http://arXiv.org/abs/1310.7915}{{\tt 1310.7915}}].

\bibitem{Kehagias:2013yd}
A.~Kehagias and A.~Riotto, {\it {Symmetries and Consistency Relations in the
  Large Scale Structure of the Universe}},  {\em Nucl.Phys.} {\bf B873} (2013)
  514--529 [\href{http://arXiv.org/abs/1302.0130}{{\tt 1302.0130}}].

\bibitem{Creminelli:2013mca}
P.~Creminelli, J.~Nore{\~n}a, M.~Simonovi{\'c} and F.~Vernizzi, {\it
  {Single-Field Consistency Relations of Large Scale Structure}},  {\em JCAP}
  {\bf 1312} (2013) 025 [\href{http://arXiv.org/abs/1309.3557}{{\tt
  1309.3557}}].

\bibitem{Rizzo:2016akm}
L.~A. Rizzo, D.~F. Mota and P.~Valageas, {\it {Nonzero density-velocity
  consistency relations for large scale structures}},  {\em Phys. Rev. Lett.}
  {\bf 117} (2016), no.~8 081301 [\href{http://arXiv.org/abs/1606.03708}{{\tt
  1606.03708}}].

\bibitem{Peloso:2015jua}
M.~Peloso, M.~Pietroni, M.~Viel and F.~Villaescusa-Navarro, {\it {The effect of
  massive neutrinos on the BAO peak}},
  \href{http://arXiv.org/abs/1505.07477}{{\tt 1505.07477}}.

\bibitem{Jain:1995kx}
B.~Jain and E.~Bertschinger, {\it {Selfsimilar evolution of cosmological
  density fluctuations}},  {\em Astrophys.J.} {\bf 456} (1996) 43
  [\href{http://arXiv.org/abs/astro-ph/9503025}{{\tt astro-ph/9503025}}].

\bibitem{Blas:2013bpa}
D.~Blas, M.~Garny and T.~Konstandin, {\it {On the non-linear scale of
  cosmological perturbation theory}},  {\em JCAP} {\bf 1309} (2013) 024
  [\href{http://arXiv.org/abs/1304.1546}{{\tt 1304.1546}}].

\bibitem{Bernardeau:2012aq}
F.~Bernardeau, N.~Van~de Rijt and F.~Vernizzi, {\it {Power spectra in the
  eikonal approximation with adiabatic and non-adiabatic modes}},
  \href{http://arXiv.org/abs/1209.3662}{{\tt 1209.3662}}.

\bibitem{Heitmann:2013bra}
K.~Heitmann, E.~Lawrence, J.~Kwan, S.~Habib and D.~Higdon, {\it {The Coyote
  Universe Extended: Precision Emulation of the Matter Power Spectrum}},  {\em
  Astrophys.J.} {\bf 780} (2014) 111
  [\href{http://arXiv.org/abs/1304.7849}{{\tt 1304.7849}}].

\bibitem{Lewis:1999bs}
A.~Lewis, A.~Challinor and A.~Lasenby, {\it {Efficient computation of CMB
  anisotropies in closed FRW models}},  {\em Astrophys.J.} {\bf 538} (2000)
  473--476 [\href{http://arXiv.org/abs/astro-ph/9911177}{{\tt
  astro-ph/9911177}}].

\bibitem{LMPR09}
J.~Lesgourgues, S.~Matarrese, M.~Pietroni and A.~Riotto, {\it {Non-linear Power
  Spectrum including Massive Neutrinos: the Time-RG Flow Approach}},  {\em
  JCAP} {\bf 0906} (2009) 017 [\href{http://arXiv.org/abs/0901.4550}{{\tt
  0901.4550}}].

\end{thebibliography}\endgroup
\end{document}